\documentclass[journal,onecolumn]{IEEEtran}

\usepackage{booktabs} 

\usepackage{multirow}
\usepackage{tabularx}
\usepackage[]{graphicx}
\graphicspath{{./figures/pdf/}{./figures/jpeg/}{./figures/jpg/}{./figures/png/}{./figures/}}
\DeclareGraphicsExtensions{.pdf,.jpeg,.jpg,.png, .jpg}

\usepackage{amsmath}
\usepackage{algorithmic}
\usepackage{array}
\usepackage{caption}
\usepackage{subcaption}
\usepackage{listings}
\usepackage{fixltx2e}
\usepackage{stfloats}
\usepackage[]{url}
\usepackage[]{hyperref}
\usepackage{listings}

\usepackage[ruled]{algorithm2e} 

\SetAlFnt{\small}
\SetAlCapFnt{\small}
\SetAlCapNameFnt{\small}
\SetAlCapHSkip{0pt}
\IncMargin{-\parindent}

\begin{document}
\title{A survey on haptic technologies for mobile augmented reality} 

\author{Carlos~Bermejo, Pan~Hui \\
			Hong Kong University of Science and Technology
		}


\IEEEtitleabstractindextext{%

\begin{abstract}
Augmented Reality (AR) and Mobile Augmented Reality (MAR) applications have gained much research and industry attention these days. The mobile nature of MAR applications limits users' interaction capabilities such as inputs, and haptic feedbacks. This survey reviews current research issues in the area of human computer interaction for MAR and haptic devices. The survey first presents human sensing capabilities and their applicability in AR applications. We classify haptic devices into two groups according to the triggered sense: \textit{cutaneous/tactile}: touch, active surfaces, and mid-air; \textit{kinesthetic}: manipulandum, grasp, and exoskeleton. Due to the mobile capabilities of MAR applications, we mainly focus our study on wearable haptic devices for each category and their AR possibilities. To conclude, we discuss the future paths that haptic feedbacks should follow for MAR applications and their challenges.

\end{abstract}


}

\maketitle
\IEEEdisplaynontitleabstractindextext

\section{Introduction} \label{sec:introduction}

AR and MAR have attracted interest from both industry and academia in the last decade. MAR enhances the real world of a mobile user with computer-generated virtual content. AR applications combine real and virtual objects in a physical environment, are interactive in real time and display an augmented view. The advances in mobile computing, computer vision, and networking have enabled the AR ecosystem.
Due to the mobile nature of MAR applications, they tend to run on mobile or wearable devices such as smartphones, tablets, smart-glasses. These device provide user's mobility, but at the cost of constrained resources such as computing-power, and energy. The computational constrains of these mobile devices limit the performance and design of AR applications in the mobile environment. Therefore, cloud infrastructures (\cite{chun2011:clonecloud},~\cite{dinh2013:survey}), computing-offloading solutions (\cite{cuervo2010:maui},~\cite{kosta2012:thinkair}), service providers, cloudlets (\cite{6834974}) and Fog computing (\cite{bonomi2012:fogIoT}) continue to deploy innovative services to provide a real time AR experience (\cite{tristan2017:futureNetworking}). Microsoft\footnote{\url{https://www.microsoft.com/en-us/hololens}}, Facebook\footnote{\url{https://developers.facebook.com/products/camera-effects/ar-studio/}} and Apple\footnote{\url{https://developer.apple.com/arkit/}} have shown their interest in AR applications and they believe in the viability of this technology. The current trend of mobile AR applications has affected the mobile market. Some well-known commercial MAR applications such as Pokemon GO\footnote{\url{http://www.pokemongo.com/}} are location-based AR mobile games. 

On the other side of the reality spectrum, Virtual Reality (VR) relates to a completely generated virtual world where the user moves, and interacts with virtual objects. VR applications will play an important role in the 
\\
future of gaming and leisure services (e.g., Facebook VR\footnote{\url{https://www.wired.com/2017/04/facebook-spaces-vr-for-your-friends/}}). These virtual scenarios bring another user experience, but the same interaction methods. Both AR and VR share similar inputs and outputs methods between the augmented/virtual reality and the user. The input methods used range from external devices such as hand controllers, wearables (e.g., Myo armband\footnote{\url{https://www.myo.com/}}), to mid-air interactions such as gaze controller or hand gesture recognition (e.g., Microsoft HoloLens). User eXperience (UX) is a key factor in these AR/VR ecosystems. An object's virtual nature needs to be supported by usable and intuitive interactions. 

In the following we can define input and output interactions. The former, corresponds to user input interactions with the environment such as hand gestures, or using an external device (e.g., hand controller, and wearables). The latter, is related to the feedbacks that the environment should provide to user actions or virtual object interactions such as sound, vibration, or visual. Although input interactions have improved during the past few years owing to the advances in computer vision, tracking, and image capturing devices such as cameras and infrared devices, the feedback provided by such environments are still primitive. Feedback appears for example as images (visual feedback), sounds (sound feedback) and/or vibration (vibrotactile feedbacks). They are aimed to provide a better UX and give the user a sense of agency (SA). The latter, SA, is related to initiating, executing and controlling a user action. As an example, users can move virtual objects with their bare hands or using a controller, but the physical characteristics of these virtual objects such as texture, size, and weight cannot be perceived without increasing the complexity of the ecosystem. There exists devices to simulate textures, and virtual objects' sizes. However, these systems require external and complicated devices to be worn by the user.  

MAR applications limit the size of haptic devices that can be used. The mobile characteristics of these environments are the focus of our related work on wearable devices. Although, there exists a myriad of developments regarding AR without size/weight limitations such active surfaces (Section~\ref{sec:tactile}) or manipulandum devices (Section~\ref{sec:kinesthetic}).

In this survey, in section~\ref{sec:haptics} we present the human sensing capabilities and their applicability to AR applications. After classifying the main groups of haptic devices regarding the involved sense, in section~\ref{sec:tactile} we present the tactile wearable devices. In Section~\ref{sec:tactile}, we also describe the three main approaches: \textit{cutaneous}, \textit{active surfaces}, and \textit{mid-air}; to provide a tactile experience. After that, we focus our study on kinesthetic feedback and the different approaches to designing the devices (Section~\ref{sec:kinesthetic}). In this Section~\ref{sec:kinesthetic} we describe the grounded devices, called manipulandum, grasping devices, exoskeleton, and other devices that do not fit within these three categories. In Section~\ref{sec:commercial}, we mention the current commercial haptic feedback devices on the market. Finally, we present the challenges and future directions for MAR haptic devices (Section~\ref{sec:challenges}) before we conclude the survey (Section~\ref{sec:conclusion}).

\section{Haptics} \label{sec:haptics}

Haptic devices enables human-computer-interaction (HCI) through touch, and external forces. Unlike traditional interfaces such as displays and sound devices, haptic devices render mechanical signals (i.e., external force) which stimulate human touch and kinesthetic channels. This field borrows from many areas, including robotics, experimental psychology, biology, computer science, system and control, and others. Due to the recent popularity of VR and AR systems, haptic devices have received great attention within the research community and entertainment industry (i.e, film, gaming, and mobile industry). The haptic ecosystem will continue its development to yield the interaction complexity needed for real-time information transmission. Visual and auditory channels are not enough to provide a perfect UX in AR/VR ecosystems. There is a need to feel (i.e., touch and move) objects in the virtual world analogously to the physical world. Haptic devices enable a bidirectional communication channel between human-computer, it creates a strong sensation of immediacy.

Haptic devices appear in multiple application scenarios~\cite{hayward2004:haptic}:
\begin{itemize}
\item \textit{Feedback reinforcement of GUIs}, such as buttons, pull-down menus
\item \textit{Games}
\item \textit{multi-media publishing}, new immersive media using VR and AR mobile platforms
\item \textit{science and data analysis}, for example in geology, data mining, multi-dimensional maps
\item \textit{arts and creation}, similar to music and painting equivalents of audio/visual communication channels
\item \textit{Sound and images edition}, help during the editing process with new user experiences
\item \textit{Vehicle industry}, better interfaces for controlling the vehicle without loosing the visual sense
\item \textit{Manufacturing}, to reduce the need of prototyping
\item \textit{Telerobotics and Teleoperation}, high quality manual controllers such as Da Vinci surgical system\footnote{\url{http://www.davincisurgery.com/}}
\item \textit{Education and training}, simulated training, and innovative passive learning methods
\item \textit{Rehabilitation}, for example the improvement of living conditions for visually impaired people
\item \textit{Scientific study of touch}
\end{itemize}


The principle operation of haptics is based on cutaneous and kinesthetic sensations. \textbf{Cutaneous/tactile}, is related to the skin; \textbf{Kinesthesia/proprioception/force}, is a sense mediated by the end sensory organs located in muscles, tendons and joints (Table~\ref{tab:haptic_devices:classification}. It is related to the capabilities of sensing the relative position of the body's parts and strength. The tactile receptors vary tremendously with the parts of the body they cover. ``Proprioceptive, or kinesthetic perception refers to the awareness of one's body state and includes the position, velocity and force supplied by the muscles''~\cite{antona2015:universal}. Together, kinesthetic and cutaneous sensations are ``fundamental to manipulation and locomotion''~\cite{arai2001:haptics} of virtual objects in VR applications.

\begin{table}%
\caption{Haptic Device Classification}
\label{tab:haptic_devices:classification}
\begin{center}
\begin{tabular}{ l l l }
  \toprule
  Group & Type & Characteristics \\
  \midrule
  \multirow{3}{*}{Cutaneous/tactile} & Cutaneous & Haptic on the user's skin (i.e., fingertip) \\
  	& Active surfaces & Communication large-scale forces and shapes and tactile information\\
    & Mid-air & Tactile feedback without contact (i.e., air, ultrasound devices) \\
  \midrule
  \multirow{3}{*}{Kinesthetic} & Manipulandum & Grounded devices with 3 to 6 Degrees of Freedom (DOF)\\
  	& Grasp & Simulates grasping interactions at the user's hand or fingers \\
  	& Exoskeleton & Grounded on the body, provide forces on natural DOF of the body \\
  \bottomrule
\end{tabular}
\end{center}
\end{table}

One of the main issues with tactile feedback is the necessary high rate refresh requirements in order to provide a satisfying experience (1kHz or more); if we compare it with video (60Hz or more) the difficulties during the design and development of cutaneous haptic devices are clear. However, the adaptability of humans also include the tactile sense and even in situations with low haptic rendering the user ignores \textbf{small} imperfections and gaps in the stimulation. Besides, the addition of other feedback channels, such as visual or audio, improves the overall haptic feedback experience. In situations where the imperfections are too obvious the realism breaks down in a similar way to videos with a lower frame rate.

A complete haptic interface can include one or several actuators such as vibroactuators, manipulandum, and sensors to measure and react to user's interactions. Besides, the feedback simulation of virtual objects is a key topic to render realistic haptic sensations. For example, the rendering of a virtual texture such as roughness for the user's cutaneous sense, or the weight of a virtual object for the kinesthetic sense. Besides, the combination of haptic feedbacks can improve the overall experience as it achieves the most realistic scenario.

\textit{Just noticeable difference (JND)} known as the \textbf{Weber-Fechner Law} ``is a measure of the minimum difference between two stimuli which are necessary in order for the difference to be distinguishable''. The JND measure is widely used in the haptic feedback ecosystem, as the cutaneous and force (kinesthetic) perceptions can differ between users and it is not easy to quantify the differences. For example in ~\cite{allin2002:measuring}, the authors use JNDs to locate thresholds of force sensitivity in virtual environments. They use a PHANToM device for their experimental study to quantify the JND for force feedback.

Furthermore, the developments in wearable haptic devices open new challenges and innovative ideas for VR, AR and the forthcoming \textit{tactile Internet}~\cite{Fettweis2014}. The high requirements regarding latency, create new topics and research paths in the networking, and sensor (i.e., electronics) fields, see more in Section~\ref{sec:challenges}.

Wearable finger-based cutaneous haptic devices are increasing in popularity as the most suitable approach for MAR/VR applications. However, there is still the need to study the detection thresholds of these devices. The insights of these studies can improve the resolution of the haptic feedbacks and reduce the network transmission of unnecessary feedback information (i.e., levels not perceived by users).

In~\cite{Lee2015}, the authors aim to quantitatively answer the JND in finger-tracking systems and the visual proprioceptive conflicts that can arise in these scenarios. In finger-tracking systems, the user interacts with the virtual environment using their fingers. The authors focus their experimental study on cutaneous finger haptic devices and a virtual environment which is rendered in a HMD (i.e., Oculus Rift\footnote{\url{https://www.oculus.com/rift/}}). The results show a JND value of 5.236\textit{cm}, so users can distinguish between their own finger and a false (i.e., rendered) one with error lager than previous JND value. Regarding the haptic feedback, the experiment presents that in small error, proprioceptive and visual situations the participant's rely on the haptic cues. In situations where the visual proprioceptive error is high, the haptic cues lose their role in the tracking system, as the users will not rely on them. The mentioned error threshold is around 3.64\textit{cm}. This paper demonstrates the improvements in finger-tracking systems using cutaneous haptic feedbacks with visual proprioceptive low error.

As within visual~\cite{naz2004:relationship} and auditive feedback~\cite{madsen1997:emotional}, tactile sensations produces different emotional responses. In~\cite{yoo2015:emotional}, the authors study the emotional responses of tactile icons. They evaluate the different responses according to four physical parameters of tactile feedback: amplitude, frequency, duration, and envelope. The authors use the valence-arousal (V-A) space to quantify the emotions~\cite{posner2005:circumplex}. The paper presents the response to tactile feedback using a vibrotactile actuator attached to the back of a smartphone. The results show that amplitude (i.e., perceived intensity) and duration have similar effects on emotional responses. Besides, the carrier frequency is strongly related to positive valence. For example more amplitude or duration corresponds with lower valence, and the greater the carrier frequency and amplitude the higher the valence and vice-versa. These insights show the relations between vibrotactile physical parameters and emotional responses. Therefore, tactile feedback needs to consider the emotional responses that can arise. In~\cite{Papetti2015}, Papetti et al. study the effect of pressing the fingertip while vibrating. Many works have focused on vibrotactile thresholds without considering the effects of applied forces in the tactile sensation. One key insight from the experiment results is that vibrotactile sensitivity depends on by applying pressure with the fingertips. Therefore, the active forces should be taken into consideration when designing vibrotactile feedback approaches.  In~\cite{Azadi2014}, the authors illustrate with several experiments how the dynamics of vibrotactile actuators change as a function of the body attachment. The authors focus on two commonly used vibrotactile displays for their experiments. The experiment results demonstrate the relation between different load forces and the vibration feedback. As with many wearable tactile displays, the tactors attached to the skin can vary the user's perception (i.e., attenuation) depending on their location.

\subsection{Other Non-Haptic Feedback} \label{subsec:other_non_haptics}

Although this paper focuses on haptic feedback and their wearability for MAR ecosystems, we need to mention other important non-haptic feedbacks such as visual and audio. The use of these non-haptics to enhance the UX is demonstrated in several papers~\cite{keele1968:processing}~\cite{pacchierotti2014:improving}, and they form the baseline case for feedback communication in several systems. 
For example, we can see their implementation in touchscreen virtual keyboards where the key is zoomed (visual) and there is a click-sound (audio) when the user presses a key. In the last decade, mobile phones provide cutaneous feedback in the form of vibration, but the visual and sound still remain as feedbacks. Moreover, in the VR and AR applications the addition of haptic feedback does not mean the suppression of any other audio-visual feedback channels, on the contrary the best approach is to combine the three sense channels. However, due to the different nature of sensory receptors (i.e., audio, visual, touch, and proprioception) the combination of these feedback complicate the design, feasibility and implementation of these realistic types of feedbacks.

Visual feedback and user actions have been strongly related since the first GUI computer system. The design of visual feedback to react to user's interactions is primordial to give an appropriate response to those actions. In~\cite{keele1968:processing},~\cite{pacchierotti2014:improving} the authors study the user's reaction time to visual feedback (190-260 ms). Furthermore, the visual improvements in current years not only focus on simpler and more intuitive methods but in faster and more fluent visual feedback. For example the visual latency of a touchscreen can hinder the UX, and scrolling speed can affect the user satisfaction while using the device.

We perceive actions that produce sound in our daily lives. For example, beeps resulting from a key press, or clang while dropping an object on the floor. In~\cite{navolio2016:egocentric}, the authors study the egocentric nature of the action-sound associations (i.e., \textit{gesture}-sound association). This question is very important in VR, and AR environments. The egocentric action-sound nature shows that users are able to learn that certain gestures create different types of sounds. Besides audio feedback can modify user's perception of kinesthetic force in VR, AR environments (i.e., change the virtual arm length, or movement). In~\cite{chang2005:audio}, the authors present different audio manipulation techniques for haptic media generation. They use the physical nature of sound waves to generate pleasant haptic feedback on a mobile phone. Although audio is a strong sensory feedback, the authors implement additional vibration feedback to improve the UX.

The addition of more feedback channels such audio, sound and tactility can improve the interaction performance in many situations. However, due to our strong reliance on our visual and auditory senses, there can be situations when the addition of a haptic feedback does not provide any improvement~\cite{akamatsu1995:comparison}. In this paper, the authors compare different feedback conditions: normal (no feedback), auditory, tactile and combined in a target selection of task experiments. Although, the results show that there is no significant difference in response times, error rates or bandwidth; there is a significant difference in the final positioning times (i.e., mouse cursor entering the target to selecting the target). Regarding the final positioning, the cutaneous feedback was the fastest. The combined feedback (visual, auditory and tactile) is also very similar to the tactile feedback condition. Therefore, there is not any increasing effect in combining sensory information for this particular task. However, the addition of other feedback channels is expected to yield to performance improvements, such as audio-visual together with haptic in the VR, and AR ecosystems.

\subsection{Wearable Haptic Devices} \label{subsec:wearable_haptics}

The portable and mobile nature of MAR applications impose several design limitations on the haptic feedback we can implement~\cite{chatzopoulos2017:mobile}.
These wearable devices are required to be light weight, small and comfortable. Furthermore, the device should not limit the user's movements, and their body degrees-of-freedom (DOFs).
For example, in fingertip actuators we need to consider that this design restrains the free movement of the user's finger and limits the use of the finger for other tasks such as typing on touchscreen devices. Wearable devices need to consider the body DOF in its design to avoid any restrictions on user's movement.

A wearable haptic device should follow the design principles in Table~\ref{tab:design_principles}. As a haptic device, they are required to follow the haptic device guidelines such as feedback intensity, fidelity factor, comfort, and mobility. The addition of portability includes more constrains to the wearable device design such as form factor, comfort, mobility, and autonomy.

\begin{table}%
\caption{Design Principles Wearable Devices}
\label{tab:design_principles}
\begin{center}
\begin{tabular}{ l l }

 \multicolumn{2}{c}{Wearable Devices} \\
  \toprule
  \multirow{2}{*}{Form factor} & Device size and shape according to the body part it is to be attached to. \\
   	& Depends on the actuators and feedback type (i.e., kinesthetic, tactile, mid-air) \\
  \midrule
  \multirow{3}{*}{Weight} & The weight of the device is relative to the part of the body where it is worn. \\
  		& For example, devices worn on the leg can be heavier than on the arm. \\
        & It depends on the muscular-skeletal strength \\
  \midrule
  \multirow{3}{*}{Mobility} & The mobility of the user needs to be considered in the placement on the body of the device. \\
  		& The wearable cannot limit the mobility of any user's muscles or body parts. \\
        & The device needs to feel natural. \\
  \midrule
  \multirow{3}{*}{Comfort} & The device needs to be for long periods without discomfort. \\
  		& The device has to be adaptable and the components (i.e., shapes, edges, bands) \\
        & ergonomically flexible. \\
          \midrule
   Autonomy & Wearable devices require enough autonomy for MAR application use. \\
   \toprule
   \multicolumn{2}{c}{Haptic Devices} \\
   \midrule
   \multirow{2}{*}{Feedback} & The feedback needs to be precise and accurate.\\
   	& It should display coherently the force, and cutaneous haptics involved in the interaction \\
     \midrule
   \multirow{2}{*}{Intensity} & The intensity of the feedback has to be appropriate to the situation, \\
   	& and user force and interactions (i.e., if the user is pushing the object).\\
     \midrule
   Fidelity & Realistically rendered haptic feedback (i.e., realistic virtual object weight). \\
   \midrule
   Duration & Duration according to the scenario, enough for the user to notice that a feedback has been triggered. \\
  \bottomrule
\end{tabular}
\end{center}
\end{table}

Moreover, the wearability characteristics limit the design of actuators and their energy consumption. The devices have to include small actuators that can be portable and require small amounts of energy in order to provide good autonomy.

In the following Sections we describe the different devices according to their haptic feedback features. We mention not only the wearable devices due to their size, weight and design, but also state of the art and novel devices that have had an impact on the haptic feedback field. As previously mentioned, the portable design of cutaneous devices make them one of the first approaches that a MAR designer needs to consider to provide haptic feedback to their AR ecosystem. Although, some novel approaches of mid-air and exoskeleton to reduce their size also make them also feasible wearable devices to consider for MAR haptic feedbacks design.

\section{Tactile, Cutaneous feedback} \label{sec:tactile}

In this Section we enumerate the most novel and important haptic devices that use the cutaneous sensory system to provide haptic feedback. We group the devices by user contact interaction (Table~\ref{tab:haptic_devices:classification}), as we think it makes easier for the MAR haptic feedback designer to find the most appropriate wearable device according to the feedback they want to provide. For example in virtual texture rendering scenarios, active surfaces~\ref{subsec:active_surfaces} and mid-air devices~\ref{subsec:mid_air} feature high resolution and accurate characteristic, but the design portability hinders in some of the proposed systems (i.e., first pin arrays devices).

\subsection{Cutaneous devices} \label{subsec:cutaneous_devices}

The cutaneous/tactile approach is currently one of the most used haptic feedback devices. The vibration on our smartphones or games controllers enables cutaneous perception of user's hand when we type, or crash a car in a video-game. The miniaturization and simple design of vibration motors make them a cost effective and feasible implementation haptic technique. However, vibration patterns are difficult to distinguish in many situations such as walking, and offer limited information (i.e., duration, strength and vibration pattern). Therefore, in~\cite{Poupyrev2002}, the authors propose a haptic display feedback with a piezoceramic film actuator to enhance the vibration haptic feedback. The device can be implemented on any button, display or inside a smartphone, and features very sharp and distinctive feedback. The authors proposed a better feedback response in comparison to vibrotactile actuators in 2003. The authors demonstrate via experiments the UI design benefits together with the tactile feedback in widgets' drag and down actions. Haptic feedback such as vibration can benefit the user interaction in the MAR environment as it adds another information channel to transmit sensory data such as touching an icon, and moving it.
In~\cite{Tinwala2009}, the authors present an eye-free text input approach based on auditory and tactile stimuli (i.e., vibration) from the smartphone. Their technique uses \textit{Graffiti}~\cite{mackenzie1997:immediate} as finger input text. Besides the feedback, an eyes-free mode is provided by auditory and tactile stimuli. The input approach uses vibration for unrecognized strokes and sound such as characters spoken for recognized strokes. The participants during the evaluation copy a text in their phones using the \textit{Graffiti} input technique. The experiments show better performance for the eyes-free mode against eyes-on mode. The results show that there is no significant ranked (by participants) difference between audio, tactile, and audio-tactile feedback. For specific feedback types, the design parameters are important for the users (participants). To conclude, the duration of the feedback does not play an important role for most type of information during the experiment; it just needs to be long enough to be perceived by the participants. The study demonstrates that vibration feedback does not have better performance results than just audio scenarios.

\begin{table}%
\caption{Most distinctive cutaneous haptic devices}
\label{tab:haptic_devices:tactile:cutaneous:most}
\begin{center}
\begin{tabular}{ l l l l }
  \toprule
  Type & Device & Author/Reference & Characteristics \\
  \midrule
  \multirow{2}{*}{Vibration} & Vibration Wristband & Carcedo et al.~\cite{Carcedo2016} & Three vibration motors on a wristband device\\
  	& Smartphone vibration device & Hoggan et al.~\cite{Hoggan2009} & Vibroactuator on the smartphone \\
    \midrule
  \multirow{4}{*}{Fingertip} & 3 RRS & Chinello et al.~\cite{chinello2015:design} & 3DOFs fingertip surface render device \\
  	& Hapthimble & Kim et al.~\cite{Kim2016} & Fingertip haptic thimble device for \\
    & & &  pushing virtual buttons \\
    & Haptic Thimble & Gabardi et al.~\cite{Gabardi2016} & Fingertip voicecoil actuator \\
    \midrule
  \multirow{3}{*}{Skin} & Skin wristband & Chinello et al.~\cite{Chinello2016} & Wristband skin stretch device  \\
  	& Skin displacement & Culbertson et al.~\cite{Culbertson2016} & Skin displacement device to render  \\
    & & & pull sensations \\
    \midrule
  \multirow{2}{*}{Other} & Fingersight & Hovarth et al.~\cite{Horvath2014} & Finger camera-haptic vibration device \\
  	& BrushTouch & Strasnick et al.~\cite{strasnick2017:brushtouch} & Wristband skin brushing device \\
  \bottomrule
\end{tabular}
\end{center}
\end{table}

Hoggan et al.~\cite{Hoggan2009} report their experimental insights into the relationship and benefits of audio and tactile icons to optimize information transmission. Previous research has focused on the use of different senses to provide the same information. They also claim that there are specific parameters in audio and tactile feedback to give the user warning notifications. The authors choose five different parameters in the design of the crossmodal audio and tactile design feedback. These design parameters are: texture (i.e., different audio waveforms); spatial location (i.e., two vibration tactors on the back of the device); duration (i.e., sound duration); and tempo/rate (i.e., for both sound and vibration). The experiment setup consists of a smartphone device that displays different information types and feedback designs such as confirmations (i.e., SMS received), errors, and progress updates (i.e., downloads). This study demonstrates that the dimensions of the vibration patterns~\cite{Carcedo2016}(i.e., tempo/rate, duration, strength) enable more resolution for the haptic information transmission. Although, as the previous study~\cite{Tinwala2009}, it depends on the scenario and the user's environment. 

The dominance of visual sense can sometimes restrain our perception of the relationship between haptic and other non-haptic feedback. For example, in many studies and commercial applications such as video-games. The visual feedback and vibrotactile feedback are strongly related, if you touch a wall the haptic device vibrates. However, we can link together other non-haptic feedback such as sound to enhance the experience in non visual MAR applications. 
In~\cite{Lee2013}, the authors propose a real-time audio-level algorithm for vibrotactile sensory translation. Author's implementation improves auditory-tactile feedback, and hence, enhances users immersion. The algorithm extracts loudness and roughness from audio signals and translates them to vibrotactile perceptual metrics: intensity and roughness. The experiments confirm that the proposed translation algorithm improves sound effects corresponding to appropriate vibrotactile feedback. The proposed algorithm can be implemented in mobile devices to provide an enhanced audio-tactile user experience. The designers of haptic feedback for MAR applications have to consider the limitations of simple vibration designs and their capabilities. The small and light weight design of these actuators make them the first device to consider and almost include in any MAR application to provide a haptic communication channel between a virtual objects and a user The information transmission limitations of vibrotactile actuators has been studied extensively. The nature of these cutaneous haptic devices make the transmission of high dimensionally data difficult as, they are limited to patterns such as intensity, duration and frequency. One-dimensional vibration transmission has been studied since the phantom sensations in~\cite{alles1970:information}. In this paper, the authors study the information transmission on the skin along through the location of vibration sensors. The results of the experiments show that the phantom sensation can display information with a low learning curve. In~\cite{Seo2015}, the authors extend the dimension of the vibroactuator on smartphones using four actuators to provide two-dimensional vibrotactile capabilities. The smartphone displays the vibrotactile flows to enable the dynamic phantom sensations. The 2D vibrotactile flows can display sensations in both 2D and circles in both directions. In~\cite{Carcedo2016}, the authors present an assistive multi-vibrotactile wristband that provides color information using vibration encoding for users who are colorblind. The authors study different vibration motor displacement and the best configuration based on user's perception accuracy. Besides, they analyze the best encoding vibration pattern to be easily and quickly recognizable. The three motor designs offer high accuracy, although it depends on the user's wrist size. Duration and sequence or number of pulses and sequential vibrations offer the best encoding methods. The vibration pattern dimensionality can enable better and higher bandwidth of information transmission.

\begin{figure}[t]
\centering
\begin{subfigure}{0.2\columnwidth}
	\centering
	\includegraphics[width=\columnwidth]{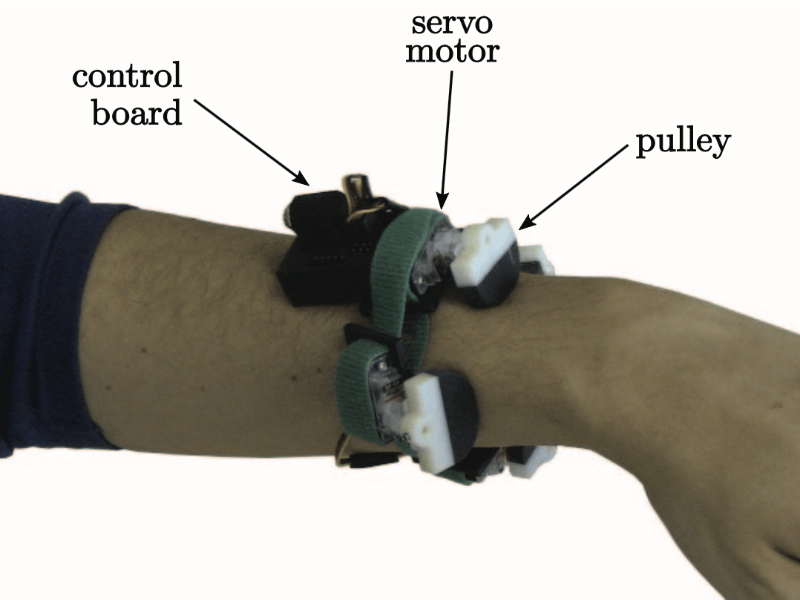}
	\caption{Skin stretch device, four actuators~\cite{Chinello2016} .}
	\label{fig:haptic:cutaneous:skin-stretch-chinello}
\end{subfigure}
\begin{subfigure}{0.2\columnwidth}
	\centering
	\includegraphics[width=\columnwidth]{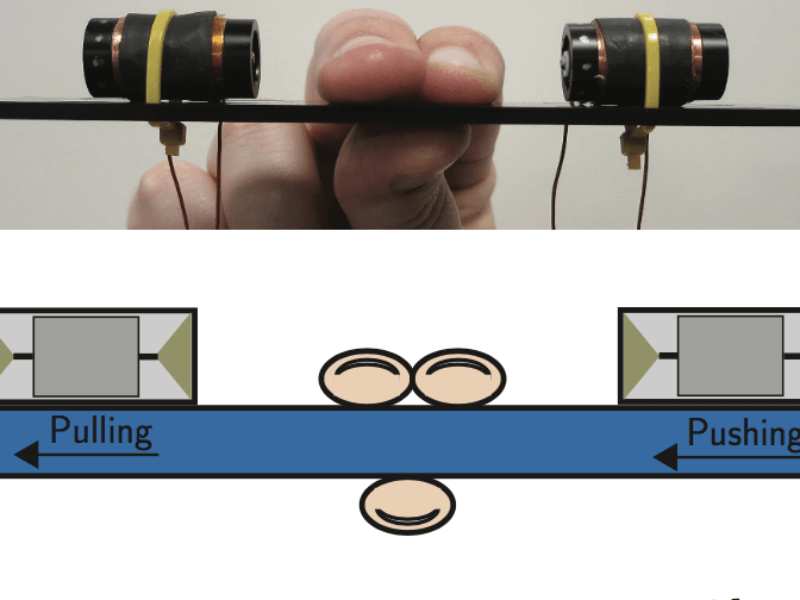}
	\caption{Asymmetric skin displacement~\cite{Culbertson2016}.}
	\label{fig:haptic:cutaneous:asymmetric-skin-displacement}
\end{subfigure}
\begin{subfigure}{0.2\columnwidth}
	\centering
	\includegraphics[width=\columnwidth]{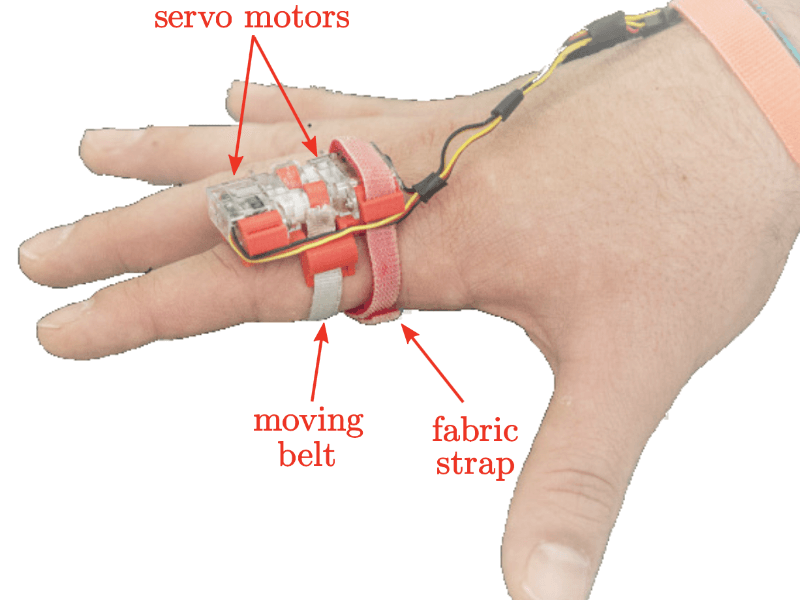}
	\caption{hRing~\cite{pacchierotti2016:hring}.}
	\label{fig:haptic:cutaneous:skin-stretch-hring}
\end{subfigure}
\begin{subfigure}{0.2\columnwidth}
	\centering
	\includegraphics[width=\columnwidth]{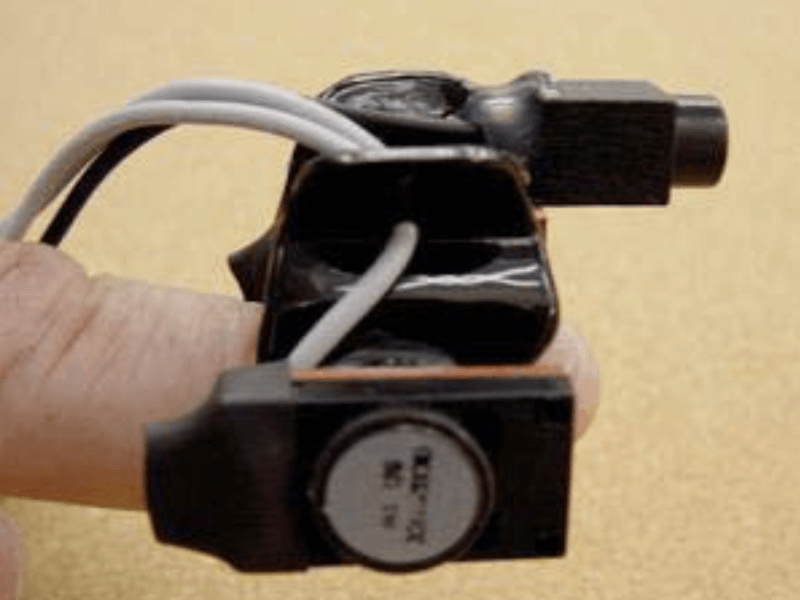}
	\caption{FingerSight, fingertip camera~\cite{Horvath2014}.}
	\label{fig:haptic:cutaneous:finger-tip-camera}
\end{subfigure}
	\caption{Wearable fingertip haptic devices. In~\ref{fig:haptic:cutaneous:asymmetric-skin-displacement}, the two voicecoil actuators simulate the asymmetric vibration on user's fingers.}
	\label{fig:haptic:cutaneous:skin} 
\end{figure}

FingerSight~\cite{Horvath2014} is a novel fingertip device for acquiring visual information through haptic channels. The visual environment information is translated into haptic stimulations. The authors' device aims to provide assistive technologies for the visually impaired. The device consists of a camera capturing device, and two speaker-based vibrators, sub-figure~\ref{fig:haptic:cutaneous:finger-tip-camera}. For the purpose of the experimental testing, the authors develop a software to detect changes in the background image color (color boundaries) so the system generates vibrations. The proposed device permits the user to scan the environment using a finger to locate specific targets (i.e., vibrates when the targeted object is found). However, the feedback interaction can be limited in 3D surroundings environments as it can not provide enough information. This wearable device can provide finger interactions with haptic feedback in MAR applications without camera-based tracking systems. Therefore, we can interact with virtual objects without the tracking technique limitations.

Currently, most communication systems provide only visual and auditive interfaces. However, gestures such as stroking, patting, and squeezing can enhance the emotional communication of current audio-visual systems. In~\cite{Rantala2014}, the authors conduct a user study to provide insights about the user's preferences for touch gestures in audio-tactile communications. The experiment setup consists of a device that can recognize user's gestures (i.e., squeezing, patting) and vibrotactile stimulation devices. Some of the participants touch the devices while others will feel the touches created on the devices (by the mentioned participants) simultaneously. Furthermore, the authors aim to map the touch gestures and their suitability during communications with a questionnaire. For example, two participants find the squeezing touch gesture as holding hands analogy. The main findings are that squeezing in audio-tactile communications and the benefit of using multiple vibrotactile actuators when the communication is only tactile is preferred. The tactile communication channel could be very useful as the future SMS in the tactile Internet (see more in Section~\ref{sec:challenges}). Kangas et al.~\cite{Kangas2014} combine gaze gesture with vibrotactile feedback to provide confirmation of interactions. In gaze gestures approaches, visual feedback can be tricky to perceive during eye movement interactions. Therefore, vibrotactile feedback can be beneficial in these situations. Authors use a gaze tracker and smartphone to enable the vibrotactile feedback. The experimental results show that haptic feedback improves efficiency of the overall gaze-experience. Sword of Elements (SoEs)~\cite{chen2016:soes} is an attachable augmented reality vibrotactile feedback device. It is an attachable solution to enhance player experience in VR environments. The device is attached to the HTC VIVE controller\footnote{\url{https://www.vive.com/us/accessory/controller/}} and features a motor module, an electronic fan (i.e., VR wind) and a thermal module.

Roudaut et al.~\cite{Roudaut2013} propose a foil overlaid touchscreen to enable spatial gesture outputs. The transparent foil device can provide up to a 1 cm motion range on a smartphone. The gesture output is non-visual and non-auditory, only tactile as it moves 
(motors for X-Y coordinates) the transparent foil along the user's finger. 
Results demonstrate that these 2-D gestures outputs are easy to learn by transfer. The novel approach of moving the touch (i.e., touchscreen) surface on the user's fingertips enables surface rendering capabilities on a portable device. We can think of different speed movements to render the moving of virtual objects in a MAR application.

HACHIStack~\cite{Hachisu2013} is a novel system which can estimate the contact time in the touchscreen of approaching objects towards a touchscreen.
This sensing method not only provide estimation of the approaching velocity, it also reduces touch latency experienced with current touch screens. In summary, HACHIStack can provide high-speed, velocity of the approaching object, and contact time prediction on a touch screen using two-layer photosensors. The authors have tried to solve the latency problem of touchscreen with a photosensor layer and a prediction algorithm, and hence render haptic feedback in real-time, without delays. 

In~\cite{Corsten2015} the authors introduce a passive back-of-device tactile landmarks to estimate finger location without seeing the screen's device. The authors propose several landmark designs and study the performance with the base case (no landmarks). The experiment results show that the back-of-device tactile landmarks outperform the base case. Hudin et al.~\cite{Hudin2015} present a system that renders independent tactile stimuli to multiple fingers without tracking their positions on a transparent surface. The tactile rendering approach is based on wave time-reversal focusing, which ``enables the spatial and temporal mechanical waves rendering using multiple transducers to create an impulse response''. The system can provide multiple focus simultaneously and therefore multitouch tactile simulation.

\begin{figure}[t]
\centering
\begin{subfigure}{0.3\columnwidth}
	\centering
	\includegraphics[width=\columnwidth]{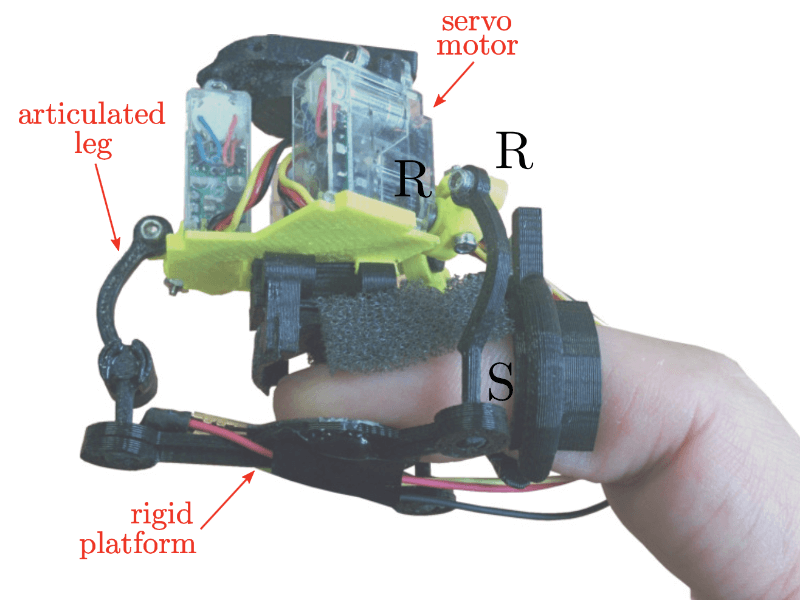}
	\caption{3-RSS~\cite{Chinello2016}.}
	\label{fig:haptic:cutaneous:finger:3rrs}
\end{subfigure}
\begin{subfigure}{0.3\columnwidth}
	\centering
	\includegraphics[width=\columnwidth]{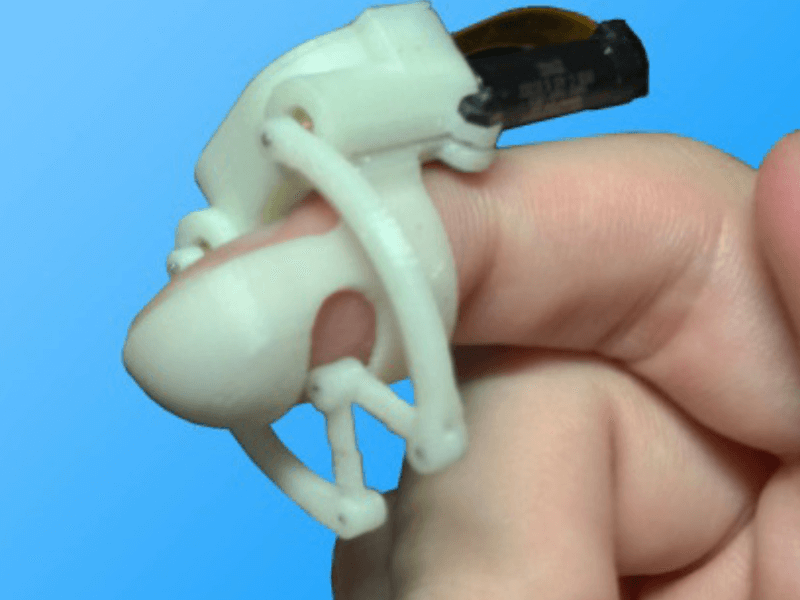}
	\caption{LinkTouch~\cite{tsetserukou2014:linktouch}.}
	\label{fig:haptic:cutaneous:finger:linktouch}
\end{subfigure}
\begin{subfigure}{0.3\columnwidth}
	\centering
	\includegraphics[width=\columnwidth]{hring-pacchierotti}
	\caption{hRing~\cite{pacchierotti2016:hring}.}
	\label{fig:haptic:cutaneous:finger:hring}
\end{subfigure}
	\caption{Wearable fingertip haptic devices.}
	\label{fig:haptic:cutaneous:finger} 
\end{figure}

Finger wearable devices have been relatively well studied in several works, as they provide a small device design and uses the user's fingertip to transmit the haptic data accurately. In~\cite{tsetserukou2014:linktouch}, the authors design a 2DOF force feedback fingertip device, (sub-figure~\ref{fig:haptic:cutaneous:finger:linktouch}). The device can represent contact conditions with a realistic directional force vector feedback. Jang et al.~\cite{Jang2014} explore the feedback possibilities for haptic cutaneous fingertip devices. Pseudo-haptic feedbacks relies on ``the visual feedback and its dominance over the tactile to present the haptic sensations''. Through experimental analysis the authors demonstrate that pseudo-haptics can render virtual stiffness by modulating visual cues. Pseudo-haptics can expand the range of perceived virtual stiffness, and if there is contradictory information between visual and haptic cues the results can be confusing for the user. Chinello et al.~\cite{chinello2015:design} present a novel 3DOF wearable cutaneous device for the fingertip (Figure~\ref{fig:haptic:cutaneous:finger:3rrs}). In comparison with other devices, the proposed in this paper enables 3DOF due to its three-leg articulated design. This design will be followed in other papers and studies as it offers good surface rendering performance in a small package. Pacchierotti et al.~\cite{Pacchierotti2015} present an innovative approach to remote cutaneous interaction methods. This approach works with any haptic fingertip device. The one proposed algorithm maps the remote sensed data to the motor or actuators inside the haptic device. The model can sense, and remotely render the haptic sensations in 3DOF scenarios. For the analysis of the model, the authors collect recorded interactions with a sensor (BioTac\footnote{\url{https://www.syntouchinc.com/sensor-technology/}}) to sense the haptic interactions and also the cutaneous feedback. BioTac mimics the sensory capabilities of a human finger. However, the algorithm has some limitations such as the vibration perception against user's applied forces, as it only renders the remote sensed data, not the current situation on the user's side. HapThimble~\cite{Kim2016} is a wearable devices which provides vibrotactile, ``pseudo-force finger sensing to mimic the physical buttons based on force-penetration for virtual screens''. HapThimble can display haptic feedback for mid-air interaction with virtual touchscreen devices. 
\textit{Haptic Thimble}~\cite{Gabardi2016} offers tactile fingertip cues in a 3DOF wearable finger device. The fingertip voicecoil actuator can rotate around a user's finger to provide a more accurate surface rendering. \textit{eShiver} is a haptic force feedback device which renders shear forces on the fingertip. \textit{eShiver} operation is very similar to \textit{ShiverPAD}~\cite{chubb2010:shiverpad}, both uses a type of electroadhesion as a friction switch. The authors use an artificial finger to measure the shear  forces in a big device setup. The simulation of friction forces in AR/MAR scenarios can provide a better understanding of the virtual object structures. However, the wearability of the proposed device is lacking. The haptic feedback not only needs to focus on the stimuli, but in the correct measure of force for different actions, and hence be able to provide accurate estimations and haptic responses. In~\cite{Mohammadi2016}, the authors propose a novel method to estimate fingertip tactile force in actions involving grasping objects. They use a custom sensing glove to estimate the contact force on the fingertip. The approach to estimating the contact force relies on magnetic, angular rate, and gravity sensors. These sensors are located along the user' phalanges. 
To study the performance of the estimation method, the authors simulate the forces are rendered them with an Omega 3 device as ground truth.

Skin displacement and stretch stimuli are another important source of cutaneous haptic devices. Furthermore, the location of these device, sometimes on the user's wrist, unlocks the fingertip mobility limitations of the finger-wearable devices. Therefore, depending on the stimuli to be rendered and MAR application scenario, these approaches can offer a better design. Chinello et al.~\cite{Chinello2016} present a novel cutaneous device capable of rendering skin stretch stimuli, sub-figure~\ref{fig:haptic:cutaneous:skin-stretch-chinello}. The device consists of four cylindrical rotating end-effectors that enable four movements on the user's wrist/arm: clockwise rotation, counter-clockwise rotation, vertical motion, and horizontal motion. The experiment demonstrates that providing skin stretch feedback benefits task completion times and errors. Furthermore, the participants find the device very useful for navigation cues. This proposed device features different skin sensations without reducing user mobility as in case of fingertip devices. In~\cite{Culbertson2016}, the authors design an asymmetric ungrounded vibration device to simulate pulling sensations through asymmetric skin displacement, sub-figure~\ref{fig:haptic:cutaneous:asymmetric-skin-displacement}. There have been previous studies that study asymmetric vibration to provide ungrounded pulling sensations. However, they have not studied the perceptual mechanism behind. The experiments show that the delay between the actuators can considerably affect the skin displacement perception. There are many studies on vibrotactile actuators, but this paper offers another skin stimuli using voicecoil actuators that can be useful in scenarios such as sliders UI, and surface texture simulations. Pacchierotti et al~\cite{pacchierotti2016:hring} claim the lack of haptic feedback approaches for wearable devices (i.e., vibrotactile sensations). This paper presents a novel wearable haptic device which consists of two servo motors and a fabric belt in contact with the user's finger skin, Figure~\ref{fig:haptic:cutaneous:skin-stretch-hring}. The presented 2DOF cutaneous device provides normal and sheer stimuli to the proximal phalanx of the user's finger. Besides, the placement of the device helps to free the user's fingertip, and the ability to interact with other hand-tracking devices such as Leap Motion\footnote{\url{https://www.leapmotion.com/}}. Bianchi et al.~\cite{Bianchi2016} present a fabric-based wearable tactile display stretching the state of a fabric that covers user's finger. The device is placed over the user's finger and attached with an elastic clip. The device mimics different stiffness levels (i.e., passive mode) and softness (i.e., active mode).

In~\cite{maisto2017:evaluation}, the authors evaluate two wearable cutaneous devices for the fingers in three different AR tasks such as hand writing, picking and moving a virtual object, and ball and cardboard (i.e., balance a virtual ball on a piece of cardboard). The first wearable device (i.e., ``3-RRS fingertip''), presented by Chinello et al~\cite{Chinello2016} uses servo motors to move a rigid platform under the fingertip. The second device, presented by Pacchierotti et al.~\cite{Pacchierotti2015} is based on a fabric belt that is controlled by two servomotors, which can stimuli friction stimuli (i.e., finger sliding over a surface) and skin stretch (i.e., both motors spin opposite directions), Figure~\ref{fig:haptic:cutaneous:finger}. The experiment setup uses an Oculus Rift HMD, camera mounted on the Oculus, and visual markers to create the AR scenario. In all of the experiments, there is a vibration feedback added to the tactile feedback provided by both devices. In the first experiment (written on a virtual board) no significant difference is seen between device performance. In the second one, grab and move virtual object, there is also no difference between wither devices. In the third experiment, balance virtual ball on cardboard, the 3-RSS device outperform the hRing. However, the hRing, due to its construction is preferred by the users as it leaves the fingerprints free for touch interactions in the physical world.

In~\cite{strasnick2017:brushtouch}, the authors propose an innovative wearable tactile stimulation feedback device through brushing. The device consists of multiple wrist-worn haptic interfaces that brush the user's skin instead of conventional vibrotactile wristband devices. The proposed device requires a greater degree of calibration in comparison with vibrotactile devices. Their experimental study shows that certain cues using brushing are better recognized than vibration.

Rekik et al.~\cite{rekik2017:localized} study the importance of surface haptic techniques to render real textures. They focus on two major approaches: Surface Haptic Object (SHO), based on the finger position; and Surface Haptic Texture (SHT), based on finger velocity. Then, they propose a new rendering technique called Localized Haptic Texture (LHT), based on the elementary tactile information that is rendered on the display, \textit{taxel}. The device to render the texture's friction is a tactile tablet which uses ultrasonic vibration to regulate the friction between the user's finger and the touchscreen. The extensive experiments demonstrate that LHT improves the quality of the tactile rendering over SHO and SHT.

Doucette et al~\cite{Doucette2013} study group interaction on digital tabletops when there is tactile feedback to people crossing their arms so as to avoid touching other people's arms. People are more aware of the management of shared public space because they try to avoid touching other people's arms. Using external devices on tabletops reduce the touching other people's arm awareness, in this paper they study the addition of vibrotactile feedback to these devices to simulate the previous scenario of avoiding touching arms. Results demonstrate that the tactile feedback improve awareness and hence the management of tabletops on public spaces. These feedback additions can be useful in MAR/AR collaborative environments. In~\cite{Obrist2015}, authors study the communication of emotions through ultrasonic haptic feedback. The tactile sense can relate to emotions such as ``anger, joy, fear, or love''. The experiment insights show that mid-air emotion communication is not arbitrary and the participants can express and recognize different tactile stimulations and associate them to a particular emotion.

\subsection{Active surfaces} \label{subsec:active_surfaces}

Active surfaces feature the best performance for rendering surfaces, with great resolution and accuracy. However, many of the aforementioned devices lack portability due to the haptic actuators design (i.e., vacuum air-based, big sized pin array actuators). 

\begin{figure}[t]
\centering
\begin{subfigure}{0.3\columnwidth}
	\centering
	\includegraphics[width=\columnwidth]{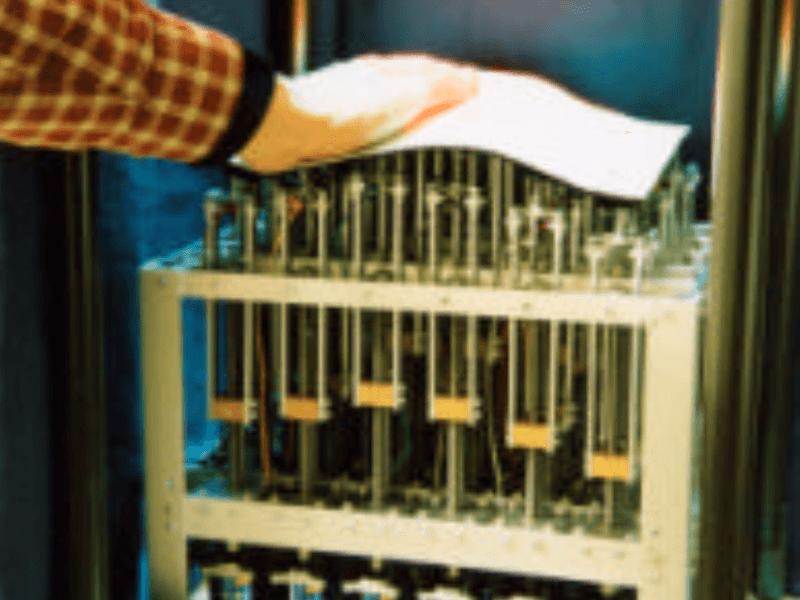}
	\caption{FEELEX~\cite{iwata2001:project}.}
	\label{fig:haptic:cutaneous:activeSurface:feelex}
\end{subfigure}
\begin{subfigure}{0.3\columnwidth}
	\centering
	\includegraphics[width=\columnwidth]{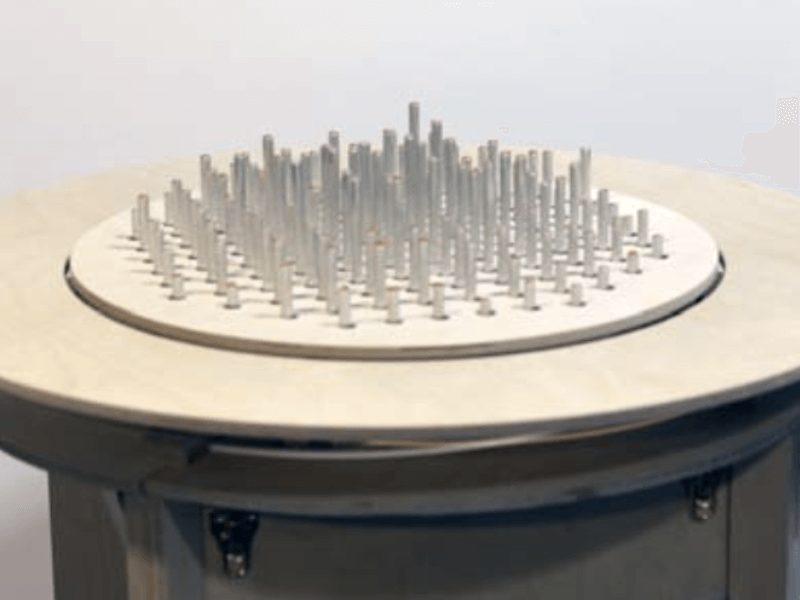}
	\caption{Pin array~\cite{leithinger2010:relief}.}
	\label{fig:haptic:cutaneous:activeSurface:arrayLeithinger}
\end{subfigure}
\begin{subfigure}{0.3\columnwidth}
	\centering
	\includegraphics[width=\columnwidth]{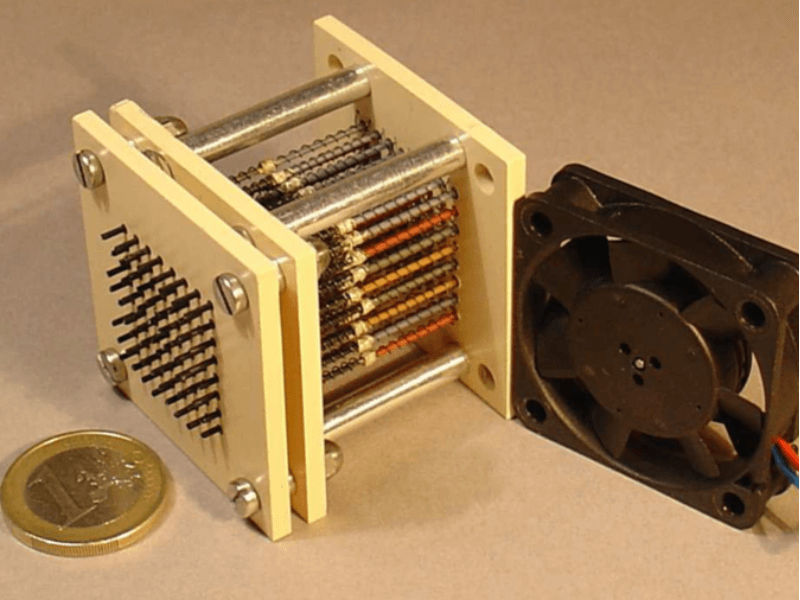}
	\caption{SMA device~\cite{velazquez2005:low}.}
	\label{fig:haptic:cutaneous:activeSurface:SMA_device}
\end{subfigure}
	\caption{Wearable fingertip haptic devices.}
	\label{fig:haptic:cutaneous:activeSurface:devices} 
\end{figure}

Iwata et al.~\cite{iwata2001:project} present FEELEX, a novel active surface haptic device that consists of an array of actuators, see 
sub-figure~\ref{fig:haptic:cutaneous:activeSurface:feelex}. The proposed device uses a matrix of linear actuators to interact with a user's palms, there is also a projector to display the GUI on the deformable screen. Although the device implementation is difficult, there are several advantages as the device provides a natural interaction with user's bare hands. Leithinger et al~\cite{leithinger2010:relief} follow a similar approach for their table top active surface design. The device consists of an array of 120 motorized pin actuators, sub-figure~\ref{fig:haptic:cutaneous:activeSurface:arrayLeithinger}. Each of the pins reacts to user input such as pulling and pushing. The authors use electric slide potentiometers as linear actuators due to their precise and fast actuation features. One of the examples of this prototype is a geospatial exploration application, whose active surface can render the landscape texture. 

\begin{table}%
\caption{Most distinctive active surfaces haptic devices}
\label{tab:haptic_devices:cutaneous:active_surfaces:most}
\begin{center}
\begin{tabular}{ l l l l }
  \toprule
  Type & Device & Author/Reference & Characteristics \\
  \midrule
  \multirow{2}{*}{Pin Array} & Pin Array &  Velazquez et al.~\cite{velazquez2005:low} & Portable pin array device to render surfaces \\
  	& Smartphone pin array & Jang et al.~\cite{Jang2016} & Pin array to deploy on smartphone sides\\
    \midrule
  \multirow{1}{*}{Multicell Array} & Multi array cell & Stanley et al.~\cite{stanley2013:haptic} & Flat deformable silicone arraycell \\
  	 \midrule
  \multirow{1}{*}{Finger-based} & Normal-/TextureTouch & Benko et al.~\cite{benko2016:normaltouch} & Fingertip device to render surfaces and textures \\
  \bottomrule
\end{tabular}
\end{center}
\end{table}

The previous devices although they are the first array of actuators prototypes, their size and weight make them unusable in mobile environments. Velazquez et al.~\cite{velazquez2005:low} presents a new low-cost, compact and light weight portable tactile display. The device consists of an array of $8\times 8$ pins based on shape memory alloy (SMA). The shape memory alloy materials are capable of recovering a predetermined shape upon heating. The device use this property on a spring to create a linear actuator, see sub-figure~\ref{fig:haptic:cutaneous:activeSurface:SMA_device}. However, owing to its nature, this material does not offer accurate control and the frequency response is affected. To heat and cool down the SMA springs the device uses electric input and a fan respectively. This wearable approach~\cite{velazquez2005:low} can be used to render surface and texture information of virtual objects on MAR applications.

\begin{figure}[t]
\centering
\begin{subfigure}{0.4\columnwidth}
	\centering
	\includegraphics[width=\columnwidth]{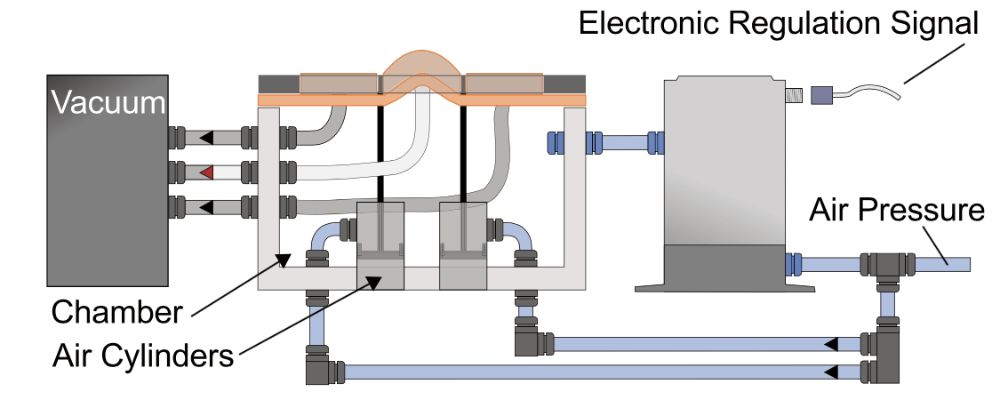}
	\caption{Particle jamming structure~\cite{stanley2013:haptic}.}
	\label{fig:haptic:cutaneous:activeSurface:particle_jamming_vacuum}
\end{subfigure}
\begin{subfigure}{0.4\columnwidth}
	\centering
	\includegraphics[width=\columnwidth]{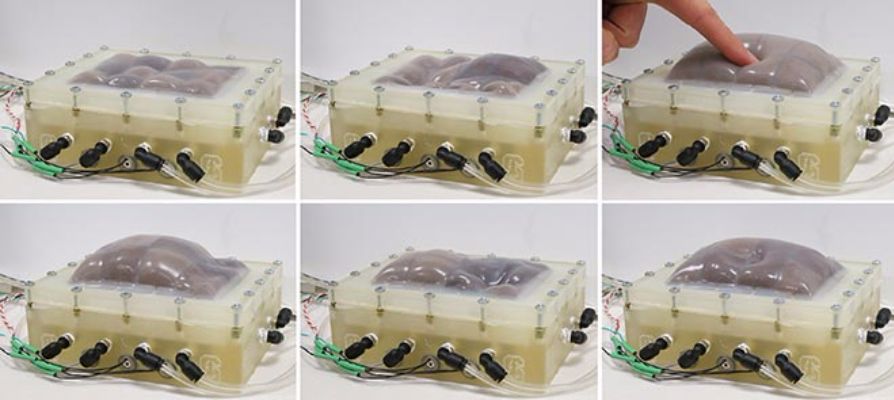}
	\caption{Particle jamming active surface~\cite{stanley2015:controllable}.}
	\label{fig:haptic:cutaneous:activeSurface:particle_jamming_device}
\end{subfigure}
	\caption{Particle jamming active surface.}
	\label{fig:haptic:cutaneous:activeSurface:particle_jamming} 
\end{figure}

Stanley et al.~\cite{stanley2013:haptic}~\cite{stanley2015:controllable} design a flat deformable layer of silicone cells regulated by air, see Figure~\ref{fig:haptic:cutaneous:activeSurface:particle_jamming}. When the air is suctioned out of the individual cells the surface changes its structure. The main difference with other active surfaces lays on the flexibility and softness of the device's surface nature, which can be very useful in medicinal training scenarios (i.e., organ rendering), or fast model prototyping.

\begin{figure}[t]
\centering
\begin{subfigure}{0.4\columnwidth}
	\centering
	\includegraphics[width=\columnwidth]{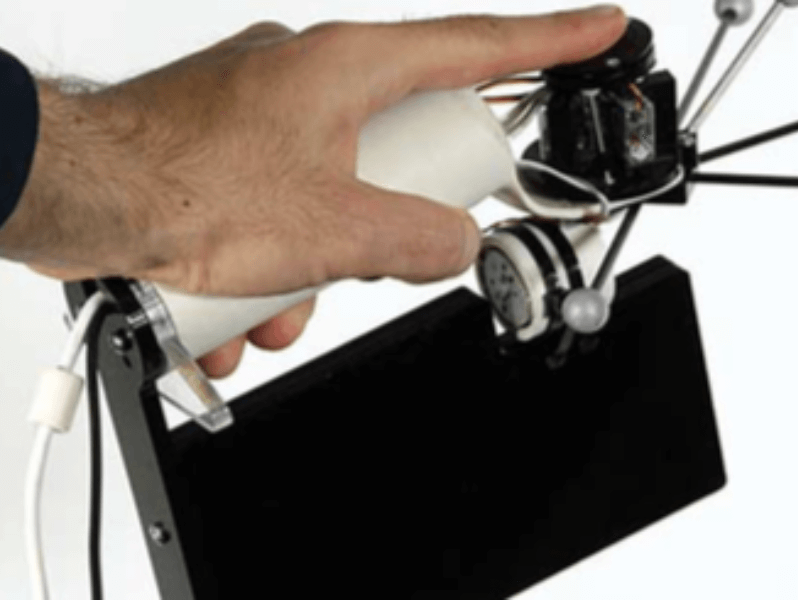}
	\caption{NormalTouch.}
	\label{fig:haptic:cutaneous:activeSurface:NormalTouch}
\end{subfigure}
\begin{subfigure}{0.4\columnwidth}
	\centering
	\includegraphics[width=\columnwidth]{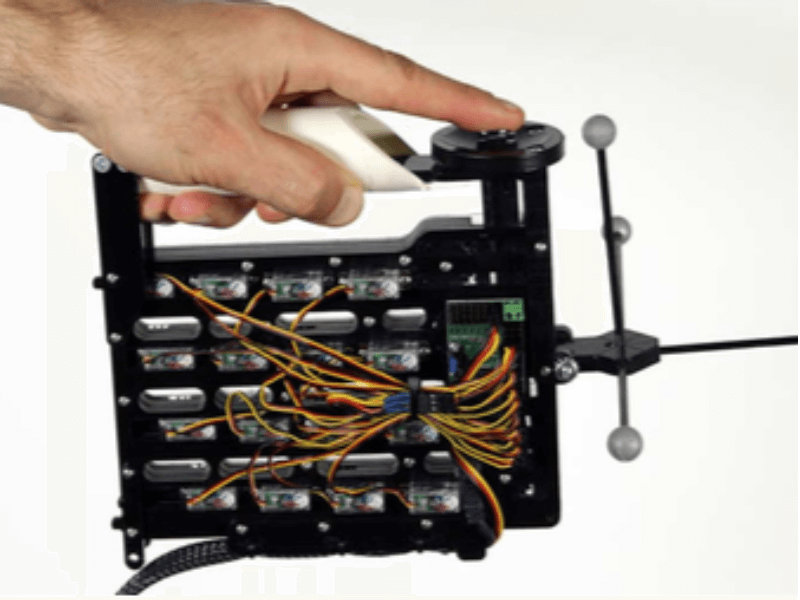}
	\caption{TextureTouch.}
	\label{fig:haptic:cutaneous:activeSurface:TextureTouch}
\end{subfigure}
	\caption{Wearable fingertip active surface devices~\cite{benko2016:normaltouch}.}
	\label{fig:haptic:cutaneous:activeSurface:benkoTouch} 
\end{figure}

In~\cite{benko2016:normaltouch}, Benko et al. propose two devices for texture rendering. The first device (\textit{NormalTouch}) consists of a handhold controller and active tiltable and extrudable platform similar to the cutaneous devices. The other proposed device (\textit{TextureTouch}), is a hand-held controller which renders 3D surfaces using a $4\times4$ array of actuate pins, see Figure~\ref{fig:haptic:cutaneous:activeSurface:benkoTouch}. \textit{NormalToch} renders the surface of the virtual object, it tilts its platform to the relative 3D (i.e., virtual object, surface) orientation and extrudes its platform according to user's controller movement. \textit{TextureTouch} works similar to the previous device. Although, it can display fine-grained surface structures. Both devices are tracked and the hand is rendered in the virtual environment. The experiment results show that both devices provide good accuracy and low error in the tracing paths and fidelity assessment tasks. The proposed devices are finger-based which imply wearability and make them another useful surface/texture of virtual objects rendering haptic devices for the MAR ecosystem. 

Jang et al.~\cite{Jang2016} propose a novel haptic edged display based on a linear touch pin array around a smartphone screen. The device not only provides an innovative haptic interaction, but can be used as a haptic notification method (i.e., changing the pin array placement). This approach can be used as feasible haptic feedback implementation for MAR applications on smartphones.

Meyer at al.~\cite{Meyer2016} present a technique to render textures on a tactile display. The authors measure a series of fingertip swipe movements across different textures and store the data as spatial friction maps. The method does not measure the velocity as its model is designed to render the texture in the spatial dimension. Due to the randomness inherent in swipes of a single texture, the authors parametrize the textures using three different distributions: Rayleigh, Rice, and Weibull. Where Weibull emerges as the most suitable distribution for fitting and categorization, the proposed method parametrizes the stochastic friction patterns in a 202-parameter model. The texture rendering is another important aspect that active surface designs need to be considered to provide a realistic UX.

\begin{figure}[t]
	\centering
	\includegraphics[width=0.5\columnwidth]{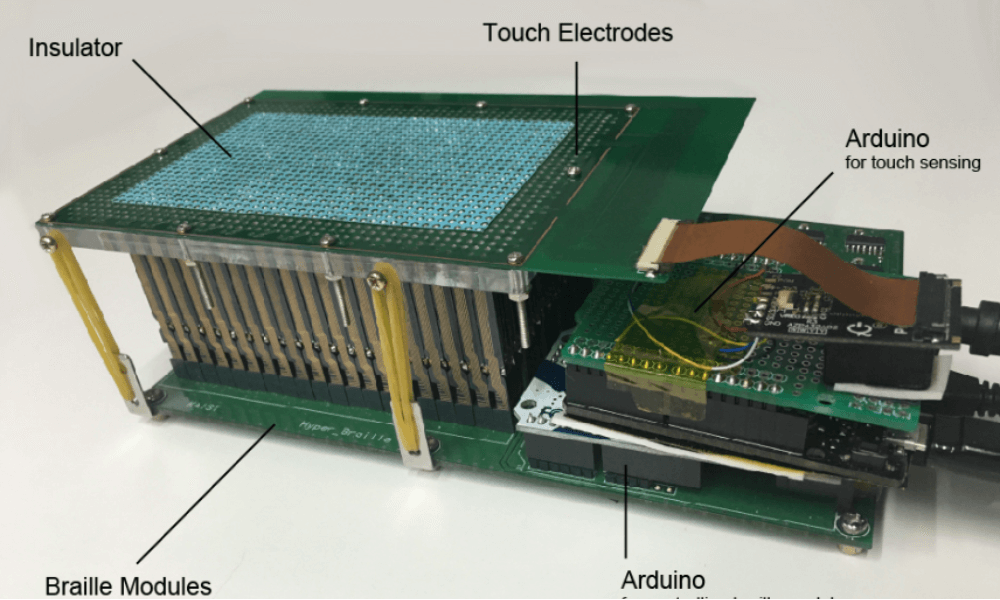}
	\caption{Pinpad, pin array device~\cite{jung2017:pinpad}.}
	\label{fig:haptic:cutaneous:activeSurface:pinPad}
\end{figure}

In~\cite{jung2017:pinpad}, authors proposed PinPad, a pin array device capable of fast and high-resolution output using a $40\times25$ array of actuated pins, Figure~\ref{fig:haptic:cutaneous:activeSurface:pinPad}. PinPad offers better spatial and temporal resolution in comparison to the state-of-art pin array devices. The experiment results show that the tactile feedback provided by PinPad enables high-resolution stimuli on the fingers. Although the prototype has some glitches in the pin movement design (i.e., stuck pin) and the noise of the device that need to be fixed in future versions. Overall, PinPad can be seen as a state-of-art pin array devices due to their high resolution and real-time speed. These pin-pad devices are used for Braille reading devices. Besides, their portability, resolution and accuracy can be included as texture/surface render devices for the MAR applications.

\subsection{Mid-air devices} \label{subsec:mid_air}

The main limitation of the previous devices corresponds to user's freewill movement that some devices apply. For example, finger-based devices constrain the use of the fingertip for other interactions such as touching a smartphone display, or as in MAR scenarios can hinder finger tracking accuracy. Mid-air devices appear as a good solution for touchless interactions, but the size and weight of many current devices do not make them a good wearable solution for MAR applications.

\begin{figure}[t]
\centering
\begin{subfigure}{0.3\columnwidth}
	\centering
	\includegraphics[width=\columnwidth]{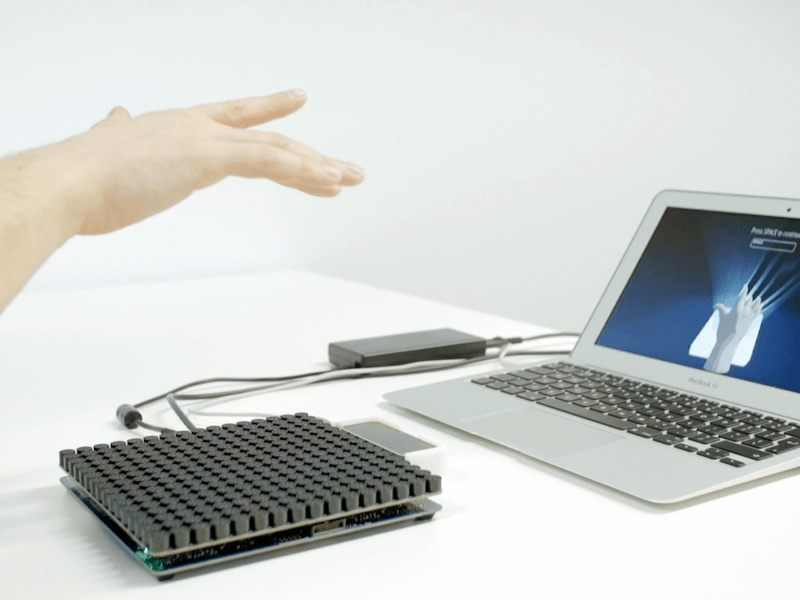}
	\caption{Ultrahaptics, ultrasound haptic~\cite{Carter2013}.}
	\label{fig:haptic:cutaneous:mid-air:ultrahaptics}
\end{subfigure}
\begin{subfigure}{0.3\columnwidth}
	\centering
	\includegraphics[width=\columnwidth]{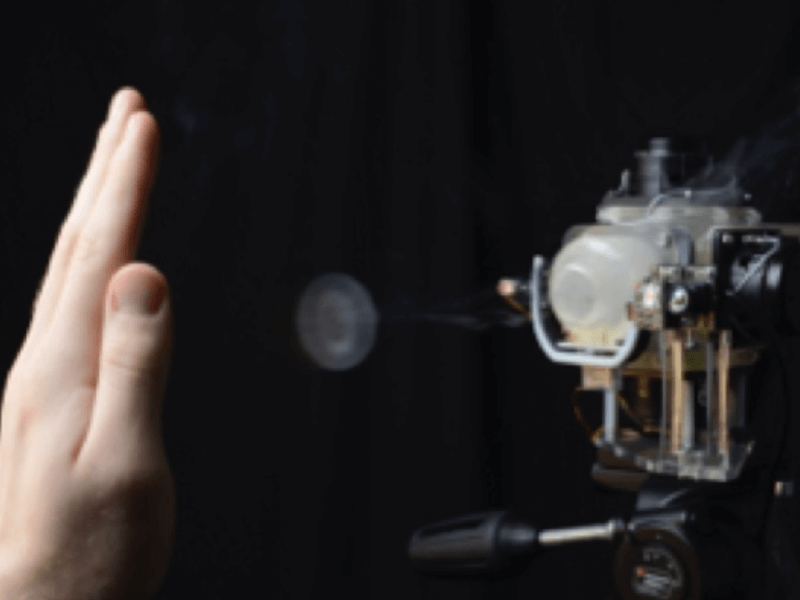}
	\caption{Aireal vortex haptic device~\cite{sodhi2013:aireal}.}
	\label{fig:haptic:cutaneous:mid-air:vortex} 
\end{subfigure}
\begin{subfigure}{0.3\columnwidth}
	\centering
	\includegraphics[width=\columnwidth]{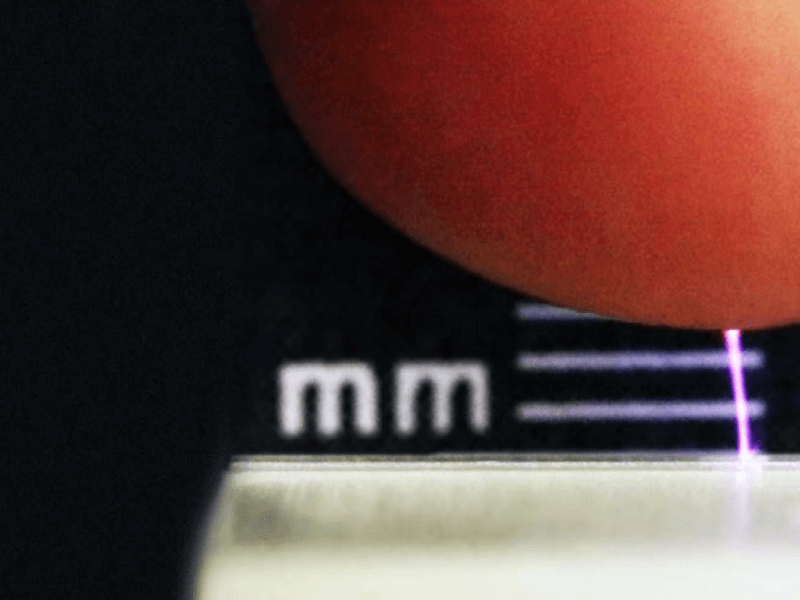}
	\caption{Electric arc array device~\cite{spelmezan2016:sparkle}.}
	\label{fig:haptic:cutaneous:mid-air:sparkle}
\end{subfigure}
	\caption{Wearable mid-air haptic devices.}
	\label{fig:haptic:cutaneous:mid-air} 
\end{figure}

Ultrasonic/ultrasound haptic devices are the most distinguishable mid-air devices and they have been studied in many related works. Moreover, due to the actuators size it is possible to wear some of these mid-air devices on the user's body. Therefore, they are a good candidate if we want to provide touchless interactions for MAR applications. Hoshi et al.~\cite{hoshi2009:adding} present an innovative hologram device with tactile sensations provided by the mid-air ultrasonic array of devices (i.e., similar devices to the commercial Ultrahaptics, sub-figure~\ref{fig:haptic:cutaneous:mid-air:ultrahaptics}). They use a mid-air display to project a floating image, a four ultrasound transducer array for the tactile display and a vision-based hand tracker system to track the user's hand. The experiment results show that the device does not provide great range and the participants have to be close to the tactile display system (i.e., transducer array). Future work includes the 3D focal point to enable tactile surface sensations moving the focal point on the user's hand. This is one of the first work with mid-air interfaces and virtual object interaction using ultrasound devices. UltraHaptics~\cite{Carter2013} is a well known and innovative mid-air ultrasonic haptic feedback. It provides multi-point haptic feedback on the user's skin. With a specific phase, amplitude, and frequency configuration the device can render different focal points and generates surfaces on the user's skin. This device will push forward many other related works on mid-air surfaces, and will stand up as one of the main mid-air haptic devices. HaptoMime~\cite{Monnai2014} enables interaction with floating images using ultrasonic tactile feedback. The device redirects the acoustic ultrasonic radiation by reflection, and hence, the ultrasonic device's beam does not interact directly with the user's skin (i.e., hand and/or finger). The device consists of a IR touch sensor frame, ultrasonic haptics, an LCD, and the Aerial Imaging Plate (AIP). The experiment results show that the tactile feedback improves user's input as it provides cues to guide the motion in mid-air (i.e., mid-air virtual keyboard). Yoshino et al.~\cite{Yoshino2013} propose a contactless touch interface with blind support using tactile interaction. The authors use a visual projector and airborne ultrasonic phased array for the tactile feedback. The audio-tactile cues help the visually impaired interact with the graphical user interface (GUI). Althought the device is not exactly wearable, the rapid advances in screen projection and ultrasonic arrays can provide future holograms or help the visually impaired to interact with AR content. SkinHaptics~\cite{Spelmezan2016} is a wearable ultrasound hand-focused device. The device consists in an ultrasound array that is attached to the user's hand to provide a tactile feedback in and through the hand. The experiment setup comprises a three by four ultrasound array matrix to simulate a numpad. One of the limitations of this feedback technique is the sensations perceived by participants on the skin and deeper inside the hand that can spread from the focus point. Makino et al.~\cite{Makino2016} propose HaptoClone, a new interactive system that enables real-time physical interactions with haptic feedbacks using ultrasonic acoustic fields and volumetric images (i.e., images displayed using micro-mirror array plates). The prototype provides limited feedback force control, and no collision detection of the occluded sides.

Air-based mid-air interfaces push air to the user's hand to render the haptic feedback sensation. Usually, these interfaces render virtual surfaces roughly as they lack good resolution and accuracy. Sodhi et al.~\cite{sodhi2013:aireal} present a novel device for mid-air tactile interaction based on air-jet approaches. Their device uses compressed air pressure to stimulate user's skin and simulate the touch sensation, sub-figure~\ref{fig:haptic:cutaneous:mid-air:vortex}. The device is able to track the user's hand (3D depth-camera) and actuate accurately on the user's skin. The device provides long distance range and easy implementation and deployment. However, due to the vortex nature of the feedback it can not provide high resolution tactile sensations. VacuumTouc~\cite{Hachisu2014} consists of a touch screen surface which sucks the air between its surface and the area where the user's finger makes contact. The authors propose several designs such as ``suction button', ``suction slider'', and ``suction dial'' that can be implemented. This paper introduces a novel haptic interface based on attractive force sensation using previously sucked air on the user's finger. In~\cite{Arafsha2015}, the authors discuss the current work on mid-air haptic feedback. Two tactile feedback methods are described in the paper: air-jet, and acoustic radiation pressure. The former uses either direct compressed air methods and vortex-based methods~\cite{sodhi2013:aireal} to simulate the tactile sensation. The latter produces tactile sensation using ultrasound~\cite{Carter2013}. The advantages of air-jet against ultrasound are its easy implementation and coverage. However, air-jet implementations have several disadvantages such as size, low spatial resolution (i.e., big focal point) and slower transfer. Both methods offer advantages but not a complete solution to interact freely with AR applications. 


\begin{table}%
\caption{Most distinctive mid-air haptic devices}
\label{tab:haptic_devices:cutaneous:active_surfaces:most}
\begin{center}
\begin{tabular}{ l l l l }
  \toprule
  Type & Device & Author/Reference & Characteristics \\
  \midrule
  \multirow{2}{*}{Ultrasound} & SkinHaptics &  Spelmezan et al.~\cite{Spelmezan2016} & Ultrasound hand haptic device \\
  & Ultrahaptics & Carter et al.~\cite{Carter2013} & Ultrasound haptic device \\
    \midrule
  \multirow{1}{*}{Air-jet} & Aireal & Sodhi et al.~\cite{sodhi2013:aireal} & Vortex-based \\
  	 \midrule
  \multirow{2}{*}{Laser-based} & LaserStroke & Lee et al.~\cite{Lee2016} & Laser device to render surface on user's palm \\
  	& & Ochiai et al.~\cite{Ochiai2016} & Laser+Ultrasound device for better \\
    & & & accuracy and haptic perception \\
  \midrule
  \multirow{1}{*}{Other} & Electric & Spelmezan et al.~\cite{spelmezan2016:sparkle} & Electric arcs for the haptic feedback \\
  \bottomrule
\end{tabular}
\end{center}
\end{table}

Laser approaches are noted owing to their accuracy and precision of the deployed systems. They are usually combined with other mid-air solutions such as ultrasound to provide the best mid-air approach. However, due to the nature of laser devices, they are dangerous to use in mobile environments, and the presented work illustrates the combination of several mid-air devices. 
LaserStroke~\cite{Lee2016} stimulates the user's tactile sense using a laser that irradiates to the user's palm, which is covered with elastic material (latex glove). The authors demonstrate the capabilities and usefulness of laser as a tactile stimulator. The thermal changes on the user's palm provide a sequence of moving tactile stimulations. LaserStroke is an interactive mid-air system that tracks the user's hand with Leap Motion and irradiates laser beams on user's palm to provide tactile stimulations. \cite{Ochiai2016} presents a new method of rendering aerial ``haptic images'' combining femtosecond-laser light fields and ultrasonic acoustic actuators. The former provides an accurate tactile perception. The latter produces a continuous and less fine-grained contact between the laser tactile perceptions. The novel combination of both mid-air fields offers the advantages of both approaches and hence, a better performance than ultrasonic or laser stimulation separately.

Spelmezan et al.~\cite{spelmezan2016:sparkle} demonstrate a novel mid-air method using touchable electric arcs on finger hover, sub-figure~\ref{fig:haptic:cutaneous:mid-air:sparkle}. To stimulate finger sensing, the device uses high-voltage arcs (safe to touch) that discharge when the finger is near the device's surface. The authors consider also the dangers and the security measures that the prototype should satisfy. The proposed device can be extended to multiple keys in the future.

Mid-air haptic devices feature the main advantage of non-covering the user's skin. Therefore, they enable many possibilities for mobility, free movement and touch experiences in the real world.

\section{Kinesthetic feedback} \label{sec:kinesthetic} 

In this Section we describe the different kinesthetic haptic approaches we can consider during the design and development of MAR applications. Kinesthetic devices display forces or motions through a tool which is usually grounded, \textit{Manipulandum} devices are not usually portable enough to consider in these scenarios. However, grasping haptic devices and exoskeletons (i.e., haptic gloves) include some wearable devices that can be used in the MAR ecosystem.

\subsection{Manipulandum} \label{subsec:manipulandum}

Manipulandum are haptic devices that render the virtual forces using grounded systems with different DOFs, depending on the characteristics of the device and/or the feedback to render.

\begin{figure}[t]
\centering
\begin{subfigure}{0.3\columnwidth}
	\centering
	\includegraphics[width=\columnwidth]{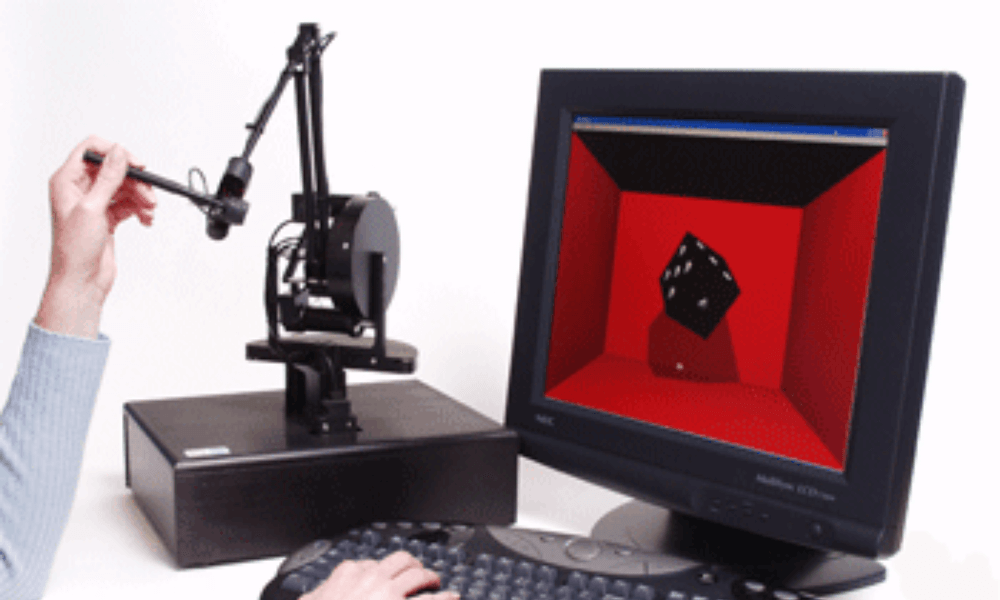}
	\caption{PHANToM~\cite{massie1994:phantom}.}
	\label{fig:haptic:kinesthetic:manipulandum:phantom}
\end{subfigure}
\begin{subfigure}{0.3\columnwidth}
	\centering
	\includegraphics[width=\columnwidth]{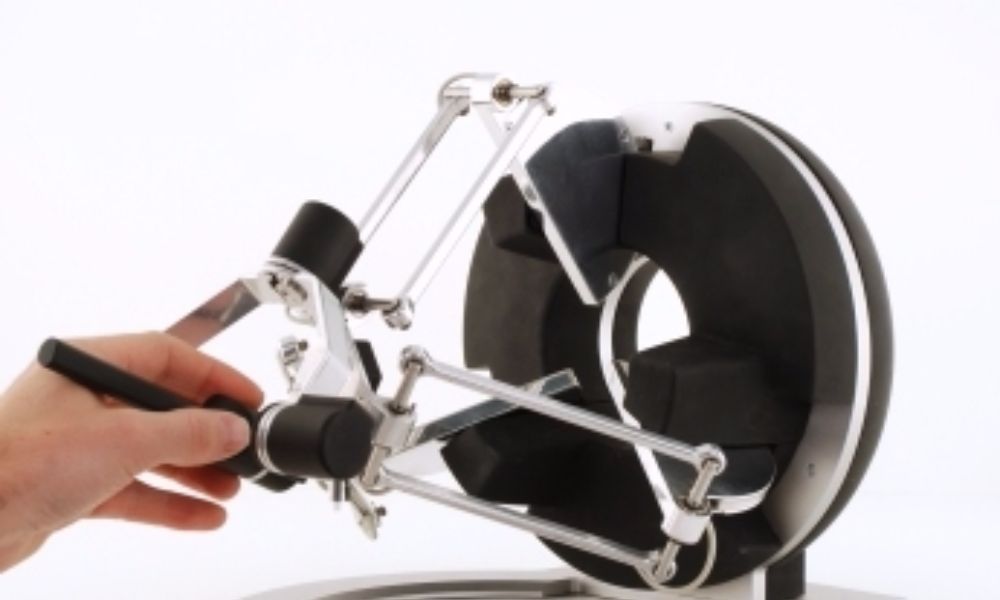}
	\caption{Omega\footnote{\url{http://www.forcedimension.com/products/omega-7/overview}}.}
	\label{fig:haptic:kinesthetic:manipulandum:omega}
\end{subfigure}
\begin{subfigure}{0.3\columnwidth}
	\centering
	\includegraphics[width=\columnwidth]{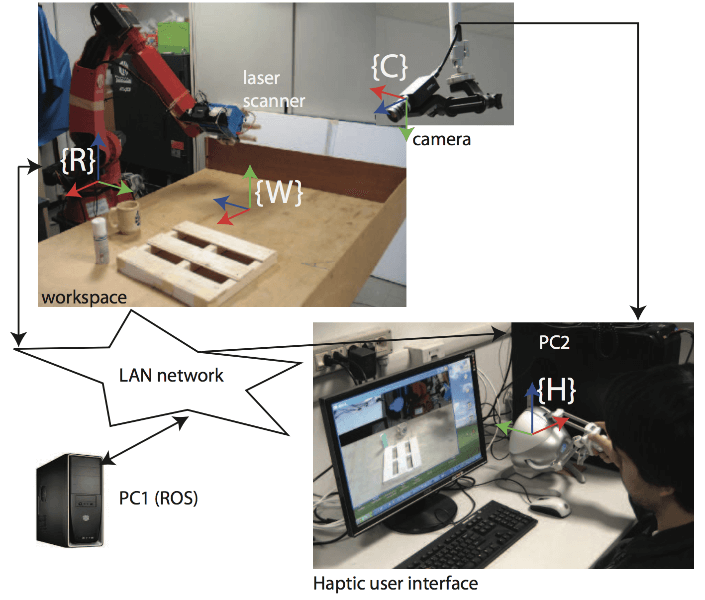}
	\caption{Robot arm~\cite{Aleotti2014}.}
	\label{fig:haptic:kinesthetic:manipulandum:robot-arm-demonstration}
\end{subfigure}
	\caption{manipulandum (kinesthetic) haptic devices.}
	\label{fig:haptic:kinesthetic:manipulandum} 
\end{figure}

In~\cite{Nojima2002}, the authors introduce a real-time AR system \textit{SmartTool}). The proposed system uses real-time sensor and haptic feedback to provide a real-time experience in comparison with other state-of-art devices at that time (2002). They use a similar device design to the PHANToM (~\cite{massie1994:phantom}) with 6DOF, sub-figure~\ref{fig:haptic:kinesthetic:manipulandum:phantom}. For real-time sensors, they have different approaches depending on the scenario such as optical sensors (i.e., cutting tissue with a scalpel), electric conductivity sensor (i.e., different fluids sensing). The use of different sensor to provide better haptic feedback is interesting in the field of AR/MAR, as most of smartphones include multiple sensors.

\begin{table}%
\caption{Most distinctive manipulandum haptic devices}
\label{tab:haptic_devices:kinesthetic:manipulandum:most}
\begin{center}
\begin{tabular}{ l l l l }
  \toprule
  Type & Device & Author/Reference & Characteristics \\
  \midrule
  \multirow{3}{*}{Grounded} & PHANToM &  Massiel et al.~\cite{massie1994:phantom} & Manipulandum haptic device \\
  & Omega & Omega Force Dimension\footnote{\url{http://www.forcedimension.com/products/omega-7/overview}} & Manipulandum haptic device \\
  & Haplet & Gallacher et al.~\cite{Gallacher2016} & Portable manipulandum \\
  \bottomrule
\end{tabular}
\end{center}
\end{table}

In~\cite{Knoerlein2007} present a collaborative ``visuo-haptic AR ping-pong game''. The system uses satellite landmarks and a hybrid tracking (IR marker based) to track the PHANToM device, which provides the haptic kinesthetic feedback. Remote collaborative scenarios, where we can share/interact with the visual-haptic feedback is still in its first deployment steps. However, it can introduce novel sharing experiences (i.e., collaborative applications with haptic feedback) in the MAR ecosystem.

In~\cite{Harders2009}, the authors propose a visio-haptic AR setup for medical simulations. They describe in detail the calibration process and their hybrid tracking techniques to provide a precise and accurate simulation environment, which is important in medical applications. For the haptic, the system uses a PHANToM device, and satellite landmarks for the tracking techniques. The system also detects occlusions of the aforementioned landmarks. The paper presents the calibration methods, hybrid tracking, synchronization and the distributed computational framework (graphics and physics distributed) to achieve less than 100 ms delay. The aforementioned delay and accuracy in AR applications is primordial to offer a good interaction between users and virtual objects.

Aleotti et al.~\cite{Aleotti2014} investigate the potential of visuo-haptic AR in programming manipulation tasks by demonstration scenarios. The system overlays a virtual environment over the physical and provides forced feedback using a 3DOF grounded device (i.e., Comau SMART SiX, similar to Omega devices), sub-figure~\ref{fig:haptic:kinesthetic:manipulandum:robot-arm-demonstration}. The system allows the user to select and manipulate virtual overlay objects. The authors also study the possibilities of the AR system in programming manipulation tasks by demonstration. The system tracks and recognizes objects in a 3D environment and the haptic interaction is achieved using a virtual proxy which interacts with virtual objects.
The system is programmed to recognized simple user's actions, the position of the objects and its manipulation. Therefore, the AR haptic environment provides not only the actions but the force the robot arm has to apply to the object during its manipulation. \textit{Haplet}~\cite{Gallacher2016} is an open-source portable haptic device which features visual, force and tactile feedback. The device has a robotic transparent arm with a vibroactuator at the tip, sub-figure~\ref{fig:haptic:other:haplet}. The device can be attached to any laptop, and smartphone display. The device is also limited in its DOF, but is an affordable device that provides good feedback for rendering textures. 

The nature of manipulandum (i.e., grounded device) makes the wearability of these devices difficult and less feasible than other approaches for MAR applications.

\subsection{Grasp devices} \label{subsec:grasp_devices}

A grasp action in the virtual world enables users interactions between user's hands and virtual objects. Then, users can push, pull, and move virtual objects as it is done in the physical world. For example, holding a glass requires a haptic device to render force or vibration on the fingertips. Furthermore, the gravity forces while holding an object can be displayed using these grasping haptic devices.

Minamizawa et al.~\cite{minamizawa2007:gravity} propose a device to present virtual object weights. The device consists of two motors that move a belt which surrounds the user's fingertip. When the two motors spin in the same direction, the belt applies a tangential force over the user's fingertip, similar to sub-figure~\ref{fig:haptic:cutaneous:finger:hring}. According to the motors' spin, the user can perceive the weight of virtual objects (i.e., two devices for each index and thumb fingers) while grasping.

\begin{table}%
\caption{Most distinctive manipulandum haptic devices}
\label{tab:haptic_devices:kinesthetic:manipulandum:most}
\begin{center}
\begin{tabular}{ l l l l }
  \toprule
  Type & Device & Author/Reference & Characteristics \\
  \midrule
    \multirow{1}{*}{Grounded} & Grasp+PHANToM &  Najdovski et al.~\cite{najdovski2014:design} &  Grounded grasping device\\
  \midrule
  \multirow{3}{*}{Finger-based} & Wolverine & Choi et al.~\cite{Choi2016} & Finger object grasping device \\
  & 3D Grasping & Handa et al.~\cite{handa2017:haptic} & 3D grasping device for 3 fingertips \\
  & Tangencial force & Minamizawa et al.~\cite{minamizawa2007:gravity} & Finger belt, skin stretch \\
  \bottomrule
\end{tabular}
\end{center}
\end{table}

\begin{figure}[t]
\centering
\begin{subfigure}{0.4\columnwidth}
	\centering
	\includegraphics[width=\columnwidth]{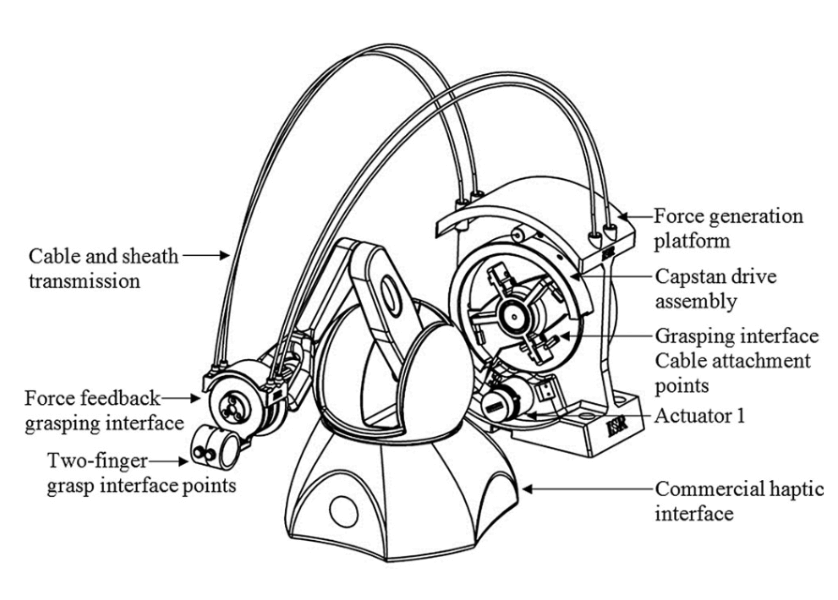}
	\caption{Pinch grasp + manipulandum device~\cite{najdovski2014:design}.}
	\label{fig:haptic:kinesthetic:grasp:pinchGrasp}
\end{subfigure}
\begin{subfigure}{0.4\columnwidth}
	\centering
	\includegraphics[width=\columnwidth]{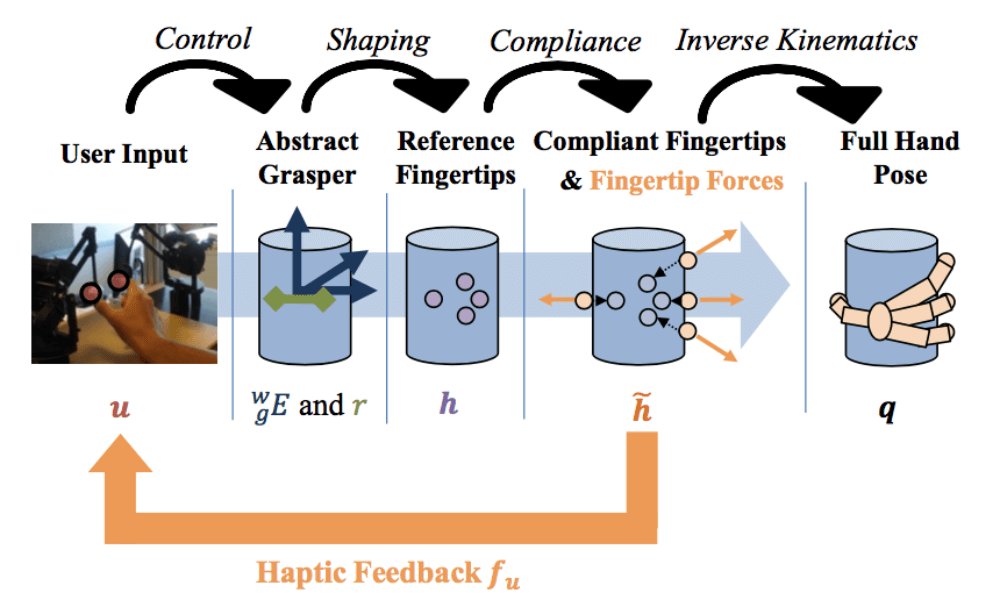}
	\caption{Hand Ons flow diagram~\cite{humberston2015:hands}.}
	\label{fig:haptic:kinesthetic:grasp:flow} 
\end{subfigure}
	\caption{Grasping devices.}
	\label{fig:haptic:kinesthetic:grasp} 
\end{figure}

In~\cite{najdovski2014:design}, the authors present an optimal configuration for a haptic device to enable pinch and grasp interactions (sub-figure~\ref{fig:haptic:kinesthetic:grasp:pinchGrasp}). The proposed device features bidirectional force on user's fingers and rotational force on the user's wrist. The pinching-grasping prototype is attached to the kinematics and force of a commercial device, Sensable Phantom Ommni\footnote{\url{http://www.geomagic.com/archives/phantom-omni/specifications/}}, which features 3DOF. The proposed device enables haptic feedback on the user's thumb and index fingers. The device's actuation is generated by a cable drive system, whose actuators are placed remotely. The system generates force on the user's finger rotation and push to render the pinch-grasp interaction. The addition of multiple feedback devices to achieve a deeper haptic experience is commonly used in grounded devices, and it is an approach that needs to be considered to enable more complex haptic feedback.

Hand Ons~\cite{humberston2015:hands} is a real-time adaptive animation interface for animating contact and precision manipulations of virtual objects with haptic feedback. The system provides contact and force information of virtual objects. Haptic feedback enhances the control and interaction with virtual objects. sub-figure~\ref{fig:haptic:kinesthetic:grasp:flow} illustrates the system steps to render a full hand pose in the virtual world with its corresponding haptic feedback. One of the limitations with the proposed system is the animation and haptic feedback changes. It is the designers/developers task to provide the corresponding cues for each scenario (i.e., object and manipulation such as grasping/holding). The authors introduce an abstraction layer the "abstract grasper" to offer more flexibility for hand animation. The hand animation is very important for providing a realistic feedback to users, the so-called visio-haptic feedback, which enhances the overall user experience in VR/AR scenarios. 

\begin{figure}[t]
\centering
\begin{subfigure}{0.4\columnwidth}
	\centering
	\includegraphics[width=\columnwidth]{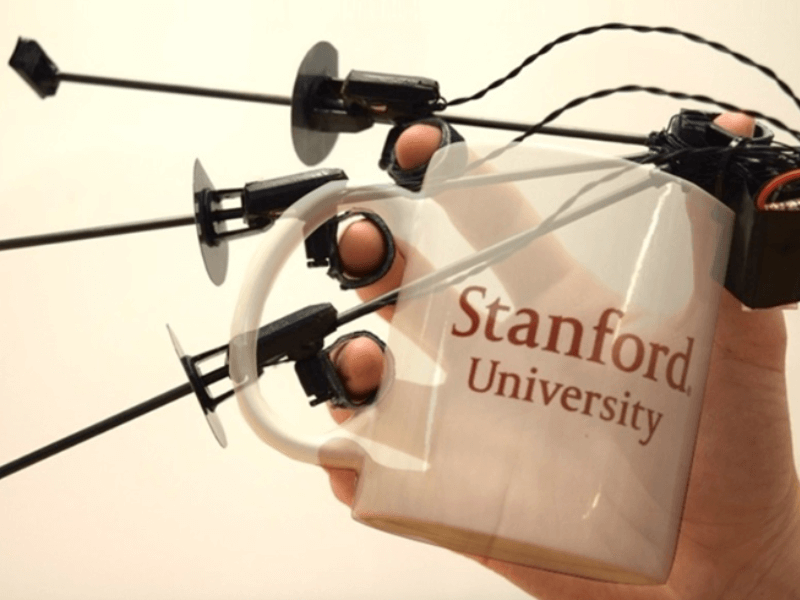}
	\caption{Wolverine~\cite{Choi2016}, grasping display of virtual objects.}
	\label{fig:haptic:kinesthetic:grasp:wolverineGrasping}
\end{subfigure}
\begin{subfigure}{0.4\columnwidth}
	\centering
	\includegraphics[width=\columnwidth]{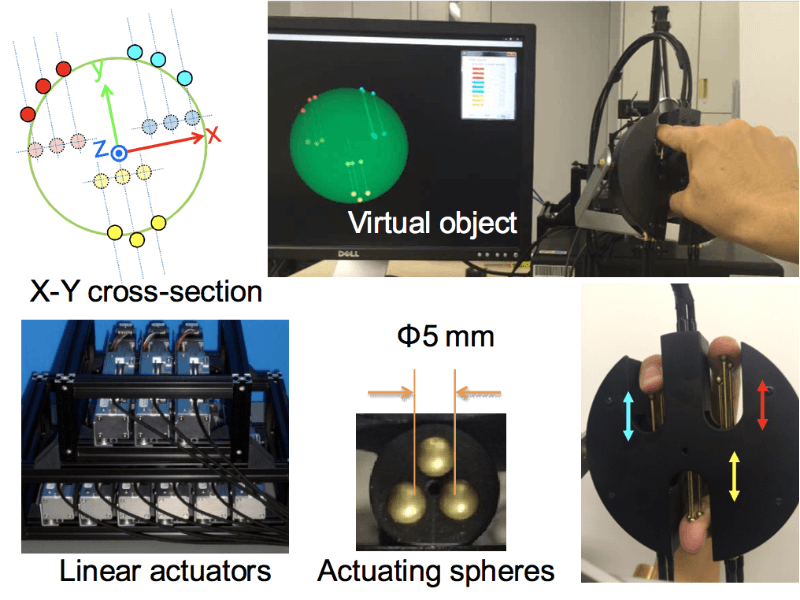}
	\caption{3D grasping device~\cite{handa2017:haptic}, linear actuators on each of the three fingers.}
	\label{fig:haptic:kinesthetic:grasp:3d_grasping}
\end{subfigure}
	\caption{Grasping devices.}
	\label{fig:haptic:kinesthetic:grasp} 
\end{figure}

Wolverine~\cite{Choi2016} is a mobile wearable haptic device designed for grasping rigid virtual objects. The authors created a light and low cost device, which renders force between the user's thumb and the three other fingers, sub-figure~\ref{fig:haptic:kinesthetic:grasp:wolverineGrasping}. The device uses low-power brake-based locking sliders to simulate the force feedback. This force mechanism renders the virtual object shape on user's hand. When users open their hand the device unlocks the braking mechanisms. In comparison with other devices such as gloves, Wolverine offers a compact design, and is an energy efficient and low cost device. 

Handa et al.~\cite{handa2017:haptic} developed a haptic display which conveys virtual objects' shapes, hardness, and textures. The proposed multi-fingered haptic 3D shape display simulates grasping actions using three fingers, sub-figure~\ref{fig:haptic:kinesthetic:grasp:3d_grasping}. The device uses external linear actuators to activate the nine spheres within the device. This paper provides an innovative device to render surfaces and simulate the user's grasping actions.

\subsection{Exoskeleton devices} \label{subsec:exoskeleton}

Exoskeleton devices are worn by a user (grounded on the body), and hence designers must decide on the importance of user's mobility while using the device and the nature of the feedback to be rendered. These haptic devices provide force on natural DOF of the body. Depending on the device feedback the size, weight and complexity could vary. For example: with grasping interfaces, haptic gloves may be sufficient; in other scenarios such as walking/running a device it may be necessary to attach a device to the legs. 

\begin{table}%
\caption{Most distinctive exoskeleton haptic devices}
\label{tab:haptic_devices:kinesthetic:exoskeleton:most}
\begin{center}
\begin{tabular}{ l l l l }
  \toprule
  Type & Device & Author/Reference & Characteristics \\
  \midrule
    \multirow{3}{*}{Gloves} & Rutgers Master II & Boutzi~\cite{bouzit2002:rutgers} &  Glove mechanical haptic device \\
    & Smart glove & Nam et al.~\cite{nam2007:smart} & Fluid-based (MR) actuators \\
  	& Jointless glove & In et al.~\cite{in2011:jointless} & Wire-based glove \\
  \bottomrule
\end{tabular}
\end{center}
\end{table}

\begin{figure}[t]
\centering
\begin{subfigure}{0.2\columnwidth}
	\centering
	\includegraphics[width=\columnwidth]{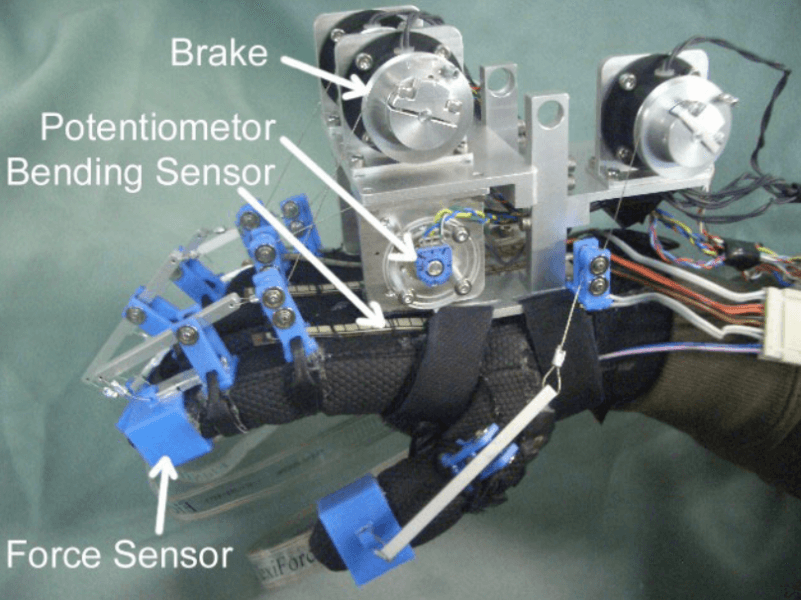}
	\caption{VR-STEF glove.}
	\label{fig:haptic:kinesthetic:exoskeleton:VR-STEF_glove}
\end{subfigure}
\begin{subfigure}{0.2\columnwidth}
	\centering
	\includegraphics[width=\columnwidth]{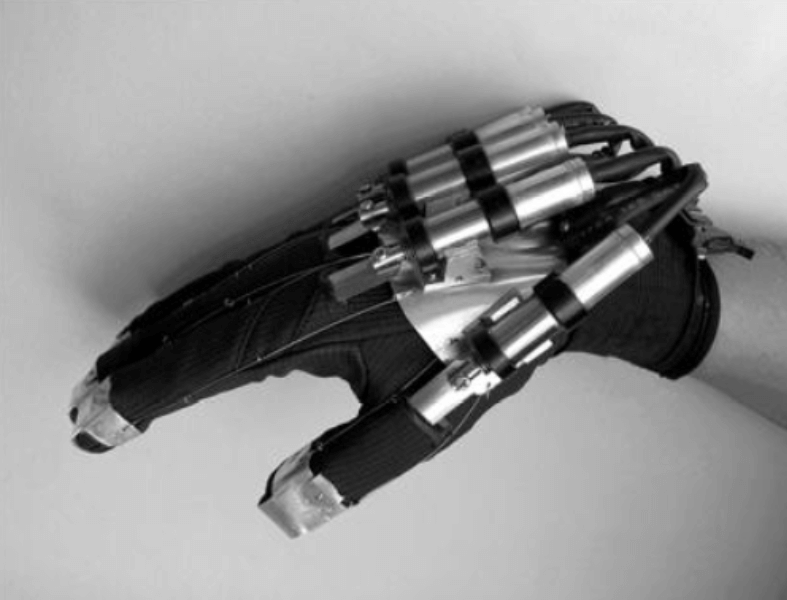}
	\caption{Step diagram.}
	\label{fig:haptic:kinesthetic:exoskeleton:MR_flex_smartGlove} 
\end{subfigure}
\begin{subfigure}{0.2\columnwidth}
	\centering
	\includegraphics[width=\columnwidth]{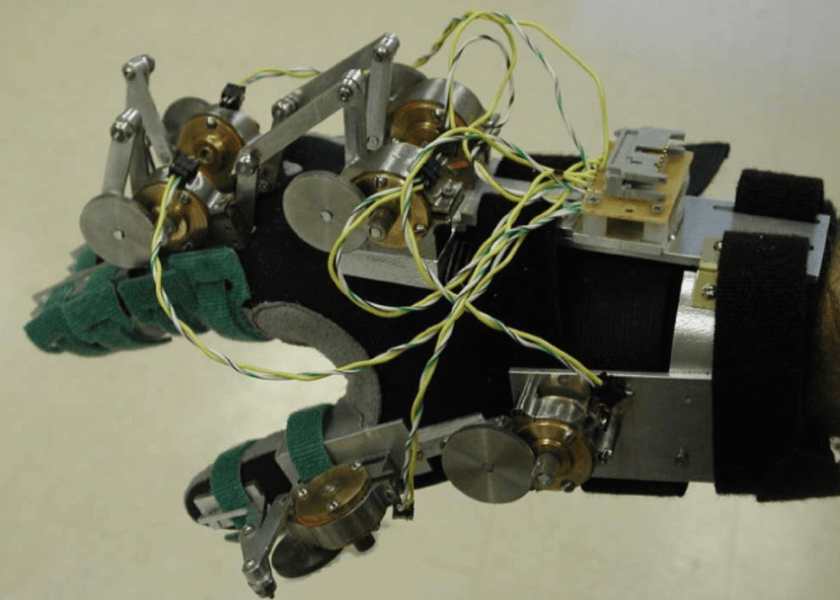}
	\caption{Step diagram.}
	\label{fig:haptic:kinesthetic:exoskeleton:MR_grasp_exoskeleton} 
\end{subfigure}
\begin{subfigure}{0.2\columnwidth}
	\centering
	\includegraphics[width=\columnwidth]{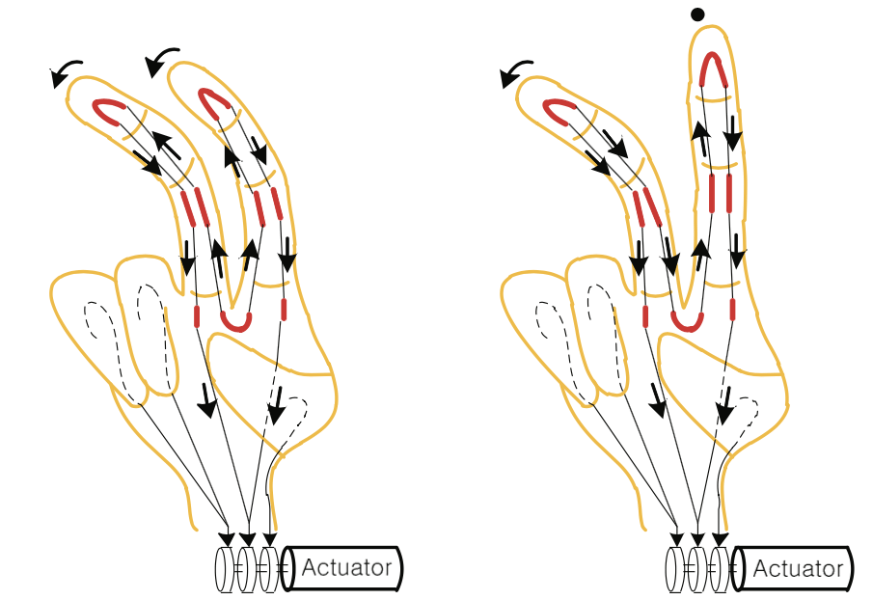}
	\caption{jointless glove.}
	\label{fig:haptic:kinesthetic:exoskeleton:jointless_glove} 
\end{subfigure}
	\caption{Grasping devices.}
	\label{fig:haptic:kinesthetic:exoskeleton} 
\end{figure}

Haptic gloves are seen as a feasible and light weight approach among exoskeleton wearable devices. They provide new force feedbacks by allowing users to pick up, grab, and feel virtual objects in a natural way. In the last decade companies have developed haptic glove devices such as \textit{Cybergrasp}\footnote{\url{http://www.cyberglovesystems.com/cybergrasp/}}, and Rutgers Master II~\cite{bouzit2002:rutgers}. One of the issues with these devices is related to the stability of haptic feedback and safety, as many of them use pneumatic/hydraulic actuators. The slow reaction time of these mechanical/fluid feedback can also hinder the overall UX. Due to the high DOF number of human hand, haptic gloves devices can focus on particular interaction approaches such as grasping, touching, pulling. In order to simplify the device design, and provide better haptic feedback, designers need to consider the haptic feedback scenarios. The actuator placement can also restrict the user's hand movement. Besides, the device design needs to be considered for mobile environments such as streets, and shops. Furthermore, the haptic interface complexity can difficult the design and implementation of the actuators, and their size and weight. In~\cite{koyanagi2005:development}, the authors present a virtual glove to recreate the ``Simple Test for Evaluation Hand Function'' (STEF). In comparison with other grounded devices such as \textit{manipulandums}, this prototype is cheaper, and simpler to setup. The rendered force is generated by electromagnetic brakes, and the force is transmitted to the fingers by wire-pulley system, see sub-figure~\ref{fig:haptic:kinesthetic:exoskeleton:VR-STEF_glove}. This passive force display glove system aims to help during the rehabilitation of stroke patients. \textit{Smart glove}~\cite{nam2007:smart} is a glove which renders haptic feedback on user's fingertips. The device uses fluid-based (magneto-rheological) actuators and a flexible link mechanisms to transmit force to the user's fingertips, see sub-figure~\ref{fig:haptic:kinesthetic:exoskeleton:MR_flex_smartGlove}. The MR fluid actuator mechanism activates when a magnetic field, by an external current, is applied. These electric activation interfaces enable faster response times. Once the  the MR fluid actuator triggers, it moves the flexible link mechanism. The device provides a better reaction time for the haptic feedback than other pneumatic/hydraulic actuators, due to the use of MR fluids. Blake et al.~\cite{blake2009:haptic} continue the MR fluid actuator approach to develop a compact haptic glove device (sub-figure~\ref{fig:haptic:kinesthetic:exoskeleton:MR_grasp_exoskeleton}). However, the device does not use a valve-type design, instead the authors opted for a disc-type design due to its portability. The device has a response time from 67 to 100 ms and a brake system (cylinder-design) 399 to 821 N$\cdot$mm torque. In~\cite{in2011:jointless}, the authors propose a jointless structure in a reduced sized for hand exoskeleton. With a jointless design the device avoids conventional pin joint structures issues and enables actuator's compactness. The device uses two wires as tendons to simulate extension and flexion, each pair of wires are embedded inside a glove (sub-figure~\ref{fig:haptic:kinesthetic:exoskeleton:jointless_glove}). The device aims to help stroke patients with their rehabilitation.

Exoskeletons in combination with other actuators such as vibration can provide haptic feedback for complex interactions. For example: grasping, holding, pushing-moving, and touching. One of the main uses of this wearable device is the reduced mobility of the user's hand for some of the prototypes mentioned, and also the hand touch isolation of the hand which is wearing the glove.


\subsection{Other kinesthetic approaches} \label{subsec:other_kinesthetic}

There are other kinesthetic approaches that use electro muscle stimulation (EMS) to display force feedback. Although, the portability and autonomy of these electronic devices are demonstrated, the rendered force lacks of continuity and can be violent in some situations. MAR application design needs to consider the features and also the constrains of EMS devices while designing an AR ecosystem.

\begin{figure}[t]
\centering
\begin{subfigure}{0.3\columnwidth}
	\centering
	\includegraphics[width=\columnwidth]{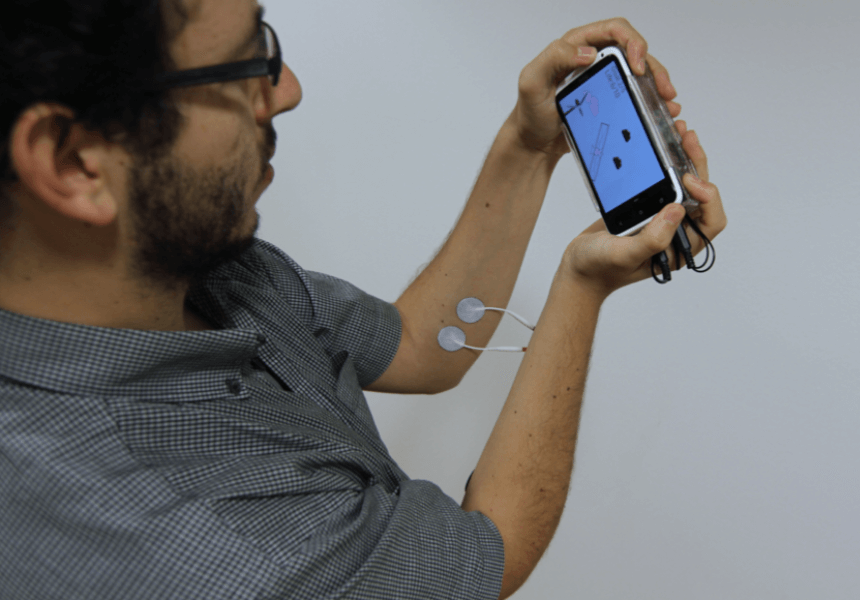}
	\caption{EMS.}
	\label{fig:haptic:kinesthetic:other:EMS_mobile}
\end{subfigure}
\begin{subfigure}{0.3\columnwidth}
	\centering
	\includegraphics[width=\columnwidth]{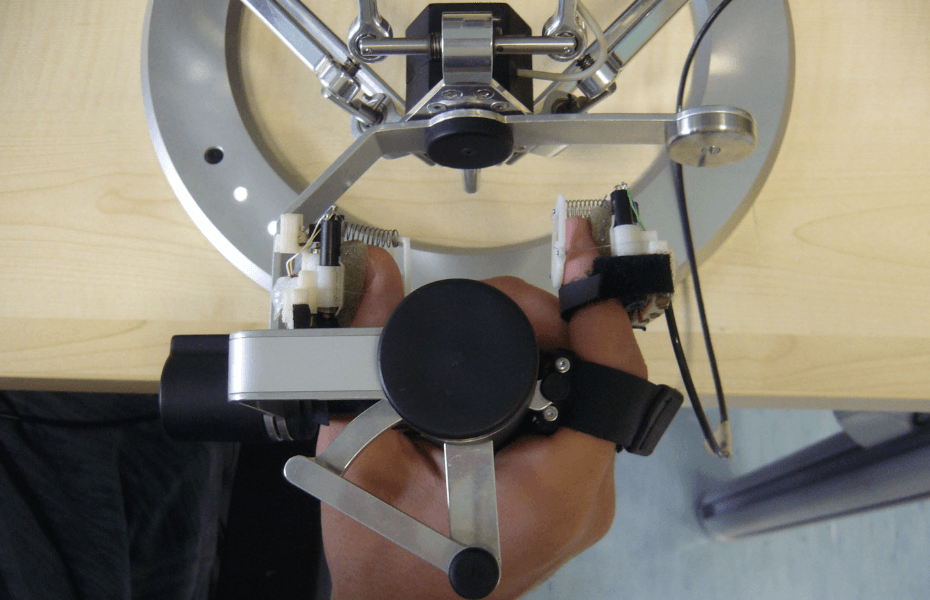}
	\caption{Sensory subtraction.}
	\label{fig:haptic:kinesthetic:other:sensorySubstraction} 
\end{subfigure}
\begin{subfigure}{0.3\columnwidth}
	\centering
	\includegraphics[width=\columnwidth]{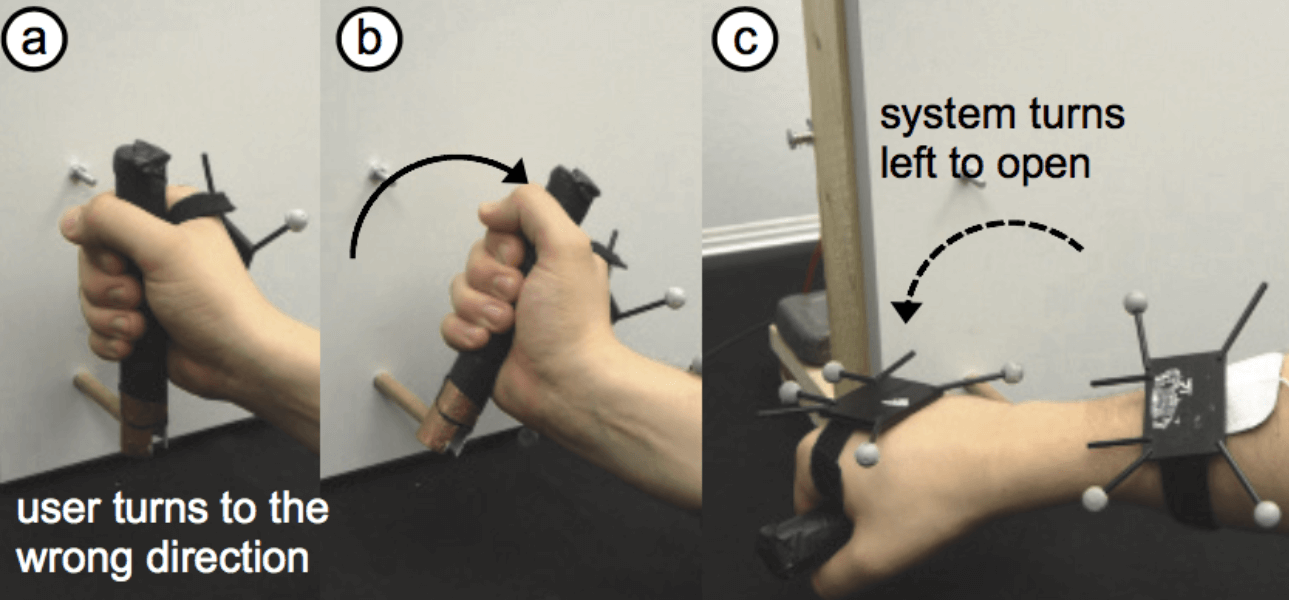}
	\caption{Affordance++.}
	\label{fig:haptic:kinesthetic:other:affordance} 
\end{subfigure}
	\caption{Grasping devices.}
	\label{fig:haptic:kinesthetic:other} 
\end{figure}

In~\cite{Lopes2013}, the authors introduced an EMS-based feedback for mobile devices. Due to  vibrotactile motor physical characteristics, the proposed system is smaller and energy efficient, sub-figure~\ref{fig:haptic:kinesthetic:other:EMS_mobile}. The authors designed a mobile game experiment to demonstrate the improved interaction with an interactive video game using this force feedback. 

Pfeiffer et al~\cite{Pfeiffer2014} compare EMS and vibration haptic feedback devices performance in free-hand interaction scenarios. They conducted several experiments to investigate the user's perception intensity for both approaches and the haptic feedback desing (i.e., vibration, EMS) to best reflect the hand-gesture. The first experiment shows that participants can differentiate in most of the cases (i.e., 80\% vibration, 90\% EMS) between two levels of intensity. The second study focus on three common gestures such as grabbing, touching, and punching and object. The results in the last experiment show that participants feel better the EMS feedback. The overall results indicate that participants feel not only better with EMS but it provides a more realistic feeling than the vibration feedback. Besides, the authors mention the privacy considerations with vibration devices, as they can be detected by others around (i.e., noise and movement), in opposition with to EMS.

In~\cite{Meli2014}, the authors present a hybrid force-cutaneous feedback approach for a robot-assisted surgery. The setup consists of a bimanual 7DOF system (i.e., provided by Omega 7\footnote{\url{http://www.forcedimension.com/products/omega-7/overview}}) and a fingerprint skin deformation device to provide tactile feedback (cutaneous devices, ~\cite{prattichizzo2013:towards},~\cite{pacchierotti2014:improving}), see sub-figure~\ref{fig:haptic:kinesthetic:other:sensorySubstraction}. Furthermore, the paper presents a technique called \textit{sensory subtraction} to substitute haptic force with cutaneous stimuli. Therefore, the device can emulate haptic force on the user's skin with both kinesthetic and cutaneous devices. The proposed device aims to outperform other sensory approaches such as substitution techniques (i.e., substituting kinesthetic forces for other haptic feedbacks). This paper demonstrates the benefits of hybrid haptic feedback approaches such as kinesthetic plus cutaneous (i.e., subtraction of haptic feedbacks instead of feedback replacement).

Impacto~\cite{lopes2015:impacto} uses a wearable device to stimulate the physical impact in VR environments. It combines tactile and EMS devices to display haptic sensations of hitting and being hit. Lopes et al.~\cite{Lopes2015a} introduce a novel concept to enhance dynamic object use. The authors propose \textit{affordance++}, an EMS device to enable object-user dynamic communication (sub-figure~\ref{fig:haptic:kinesthetic:other:affordance}). \textit{Affordance++} allows the users actuate with virtual objects, and show how the movement should be. Although EMS can be sharp and strong for mobile scenarios, the concept idea of object-behavior dynamic communication can be very useful in MAR environments.

In~\cite{Tanabe2016}, the authors describe the properties of proprioceptive sensations induced by non-grounded haptic devices. They use a vibration speaker which generates a perceived force that pulls or pushes user arms in a particular direction. This haptic asymmetric vibration design induces the sensory illusion of pulling and pushing user's hand. The experiment results show that changes in the vibroactuator input signal can alternate direction and magnitude force on users.

\section{Other haptic devices} \label{sec:other_haptic}

Lee at al.~\cite{Lee2005} propose a graspable and touchable interface based on 3D foam for AR scenarios. They use a 3D foam as a passive object that is tangible, traceable and rendered in an AR application. Therefore, the user can interact with the object in the virtual environment with overlaid information and touch/grasp the object in the physical world. The idea of passive and tangible objects to provide the haptic feedback is commonly used in VR applications. However, the usability in AR scenarios can be another feature that can improve the overall UX in MAR. For example, the addition of a simple 3D object that is rendered as another virtual object in AR application, offers not only the grasping/touching sensation but the possibility of moving a virtual-complex object on the real one.

Previous studies have focused on vibration intensity and duration along with perceived stimulus. Blum at al.~\cite{Blum2015} propose the addition of accelerometers data to improve haptic feedback of vibration actuators. For example, if the user is running the vibration will be more intense as in cases where the user is still. The addition of other surrounding information to provide better haptics according to the situation will improve the haptic feedback perception.

In~\cite{Azmandian2016}, the authors introduce a passive haptic \textit{retargeting} feedback, using dynamically aligning physical and virtual objects using our vision system (hacking human perception). The paper demonstrates three approaches that use dynamic remapping and body alignment to reuse passive haptics of same physical objects across multiple virtual objects. Although the paper focus on VR, passive haptics and dynamic remapping could be useful in the future AR/MAR applications.

\begin{figure}[t]
\centering
\begin{subfigure}{0.3\columnwidth}
	\centering
	\includegraphics[width=\columnwidth]{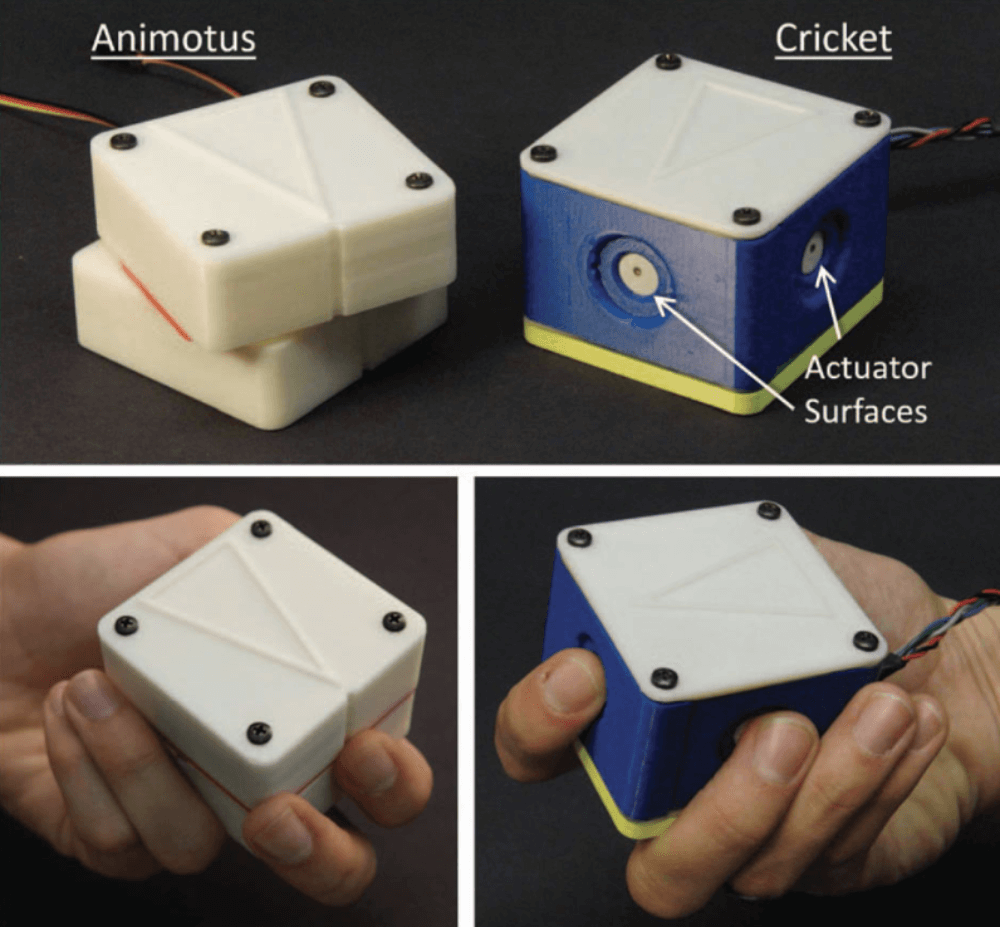}
	\caption{ Shape changing.}
	\label{fig:haptic:other:shape-changing}
\end{subfigure}
\begin{subfigure}{0.3\columnwidth}
	\centering
	\includegraphics[width=\columnwidth]{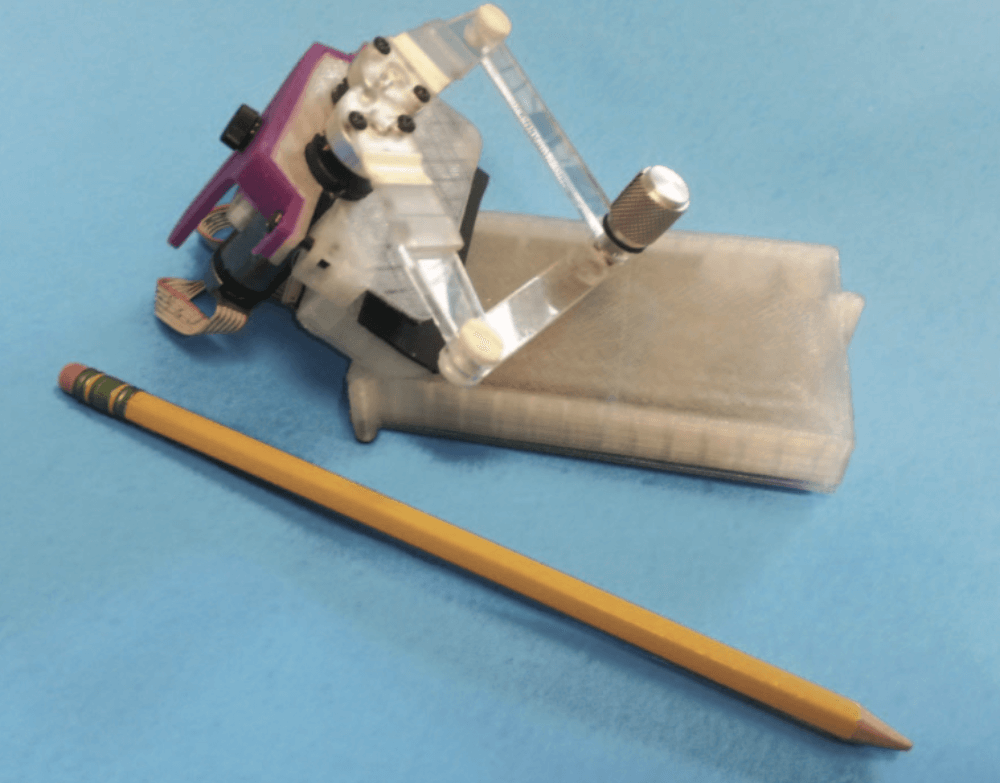}
	\caption{Haplet.}
	\label{fig:haptic:other:haplet}
\end{subfigure}
\begin{subfigure}{0.3\columnwidth}
	\centering
	\includegraphics[width=\columnwidth]{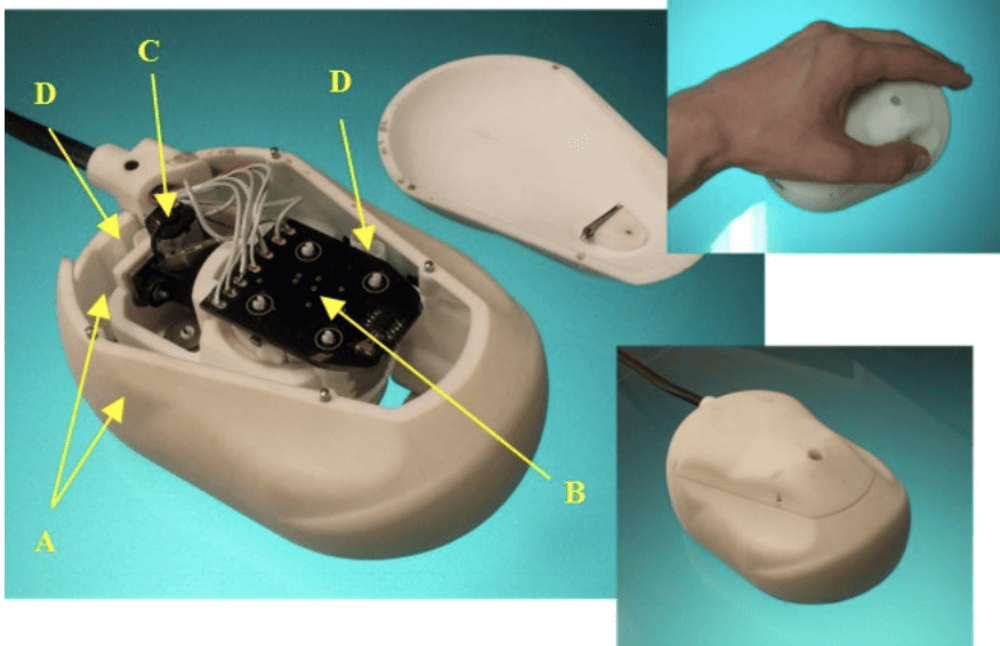}
	\caption{Mouse haptic.}
	\label{fig:haptic:other:price-mouse-haptic}
\end{subfigure}
	\caption{Other wearable haptic devices.}
	\label{fig:haptic:other} 
\end{figure}

Spiers et al.~\cite{Spiers2016} propose a shape-changing haptic interface for navigation systems. The authors compare their device with the more common vibrotactile devices used in navigation systems. The device consists of a cube shaped object with an upper half which is able to rotate and slide over its other half to display navigation cues such as forward, turn left, turn right, backwards. The experiment in an outdoor public environment shows that participants find more intuitive and pleasant the shape-changing device to navigate than the vibrotactile equivalent device, sub-figure~\ref{fig:haptic:other:shape-changing}. The addition of wearable shape-changing devices in the AR/MAR ecosystem  (i.e., textiles) enable other ways of stimuli that can be useful in no display vision situations. Furthermore, vibration feedbacks are sometimes (i.e., walking) difficult to perceive.

Price et al.~\cite{Price2016} present a touchscreen-based haptic system which features kinesthetic force feedbacks, sub-figure~\ref{fig:haptic:other:price-mouse-haptic}. The device provides static friction to simulate virtual constraints such as boundaries, area-of-effect fields and paths. The experiments demonstrate these haptic functions in a virtual maze and walls. The device aims to provide rehabilitation for upper limb stroke patients, as it engages patients with a haptic feedback. Haptic simulation of contours, boundaries and textures of virtual objects is an important topic that AR/MAR applications need to integrate in order to provide a full interaction and better UX.

Israr et el.~\cite{israr2016:stereohaptics} propose \textit{Stereohaptics}: ``a haptic interaction toolkit for tangible virtual experiences''. They aim to provide a framework and ultrasonic haptic devices to create, edit, and render rich dynamic haptic using audio-based devices.

Nakamura et al.~\cite{Nakamura2016} propose a contact pad that can emulate the sensation of softness. The haptic display provides lateral force feedback and softness rendering with electroadhesion using contact pads on the screens. The device also improves UX on displays, but it is limited to pushing and lateral force feedback.

\begin{figure}[t]
\centering
\begin{subfigure}{0.3\columnwidth}
	\centering
	\includegraphics[width=\columnwidth]{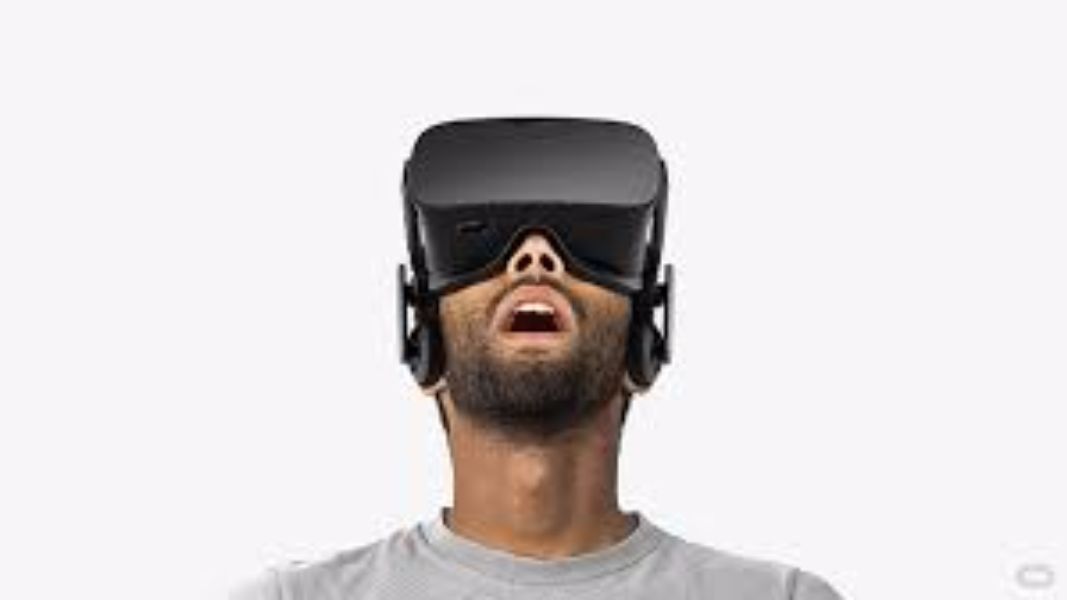}
	\caption{Oculus Rift\footnote{\url{https://www.oculus.com/rift/}}.}
	\label{fig:commercial:oculus-rift}
\end{subfigure}
\begin{subfigure}{0.3\columnwidth}
	\centering
	\includegraphics[width=\columnwidth]{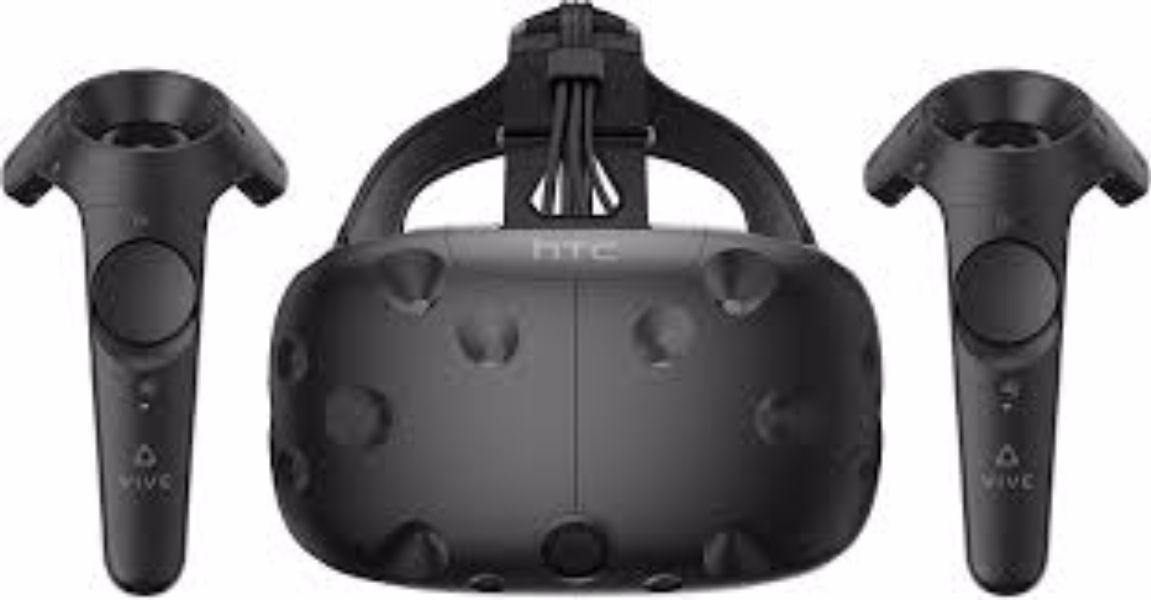}
	\caption{HTC Vive\footnote{\url{https://www.vive.com/de/}}.}
	\label{fig:commercial:htc-vibe}
\end{subfigure}
\begin{subfigure}{0.3\columnwidth}
	\centering
	\includegraphics[width=\columnwidth]{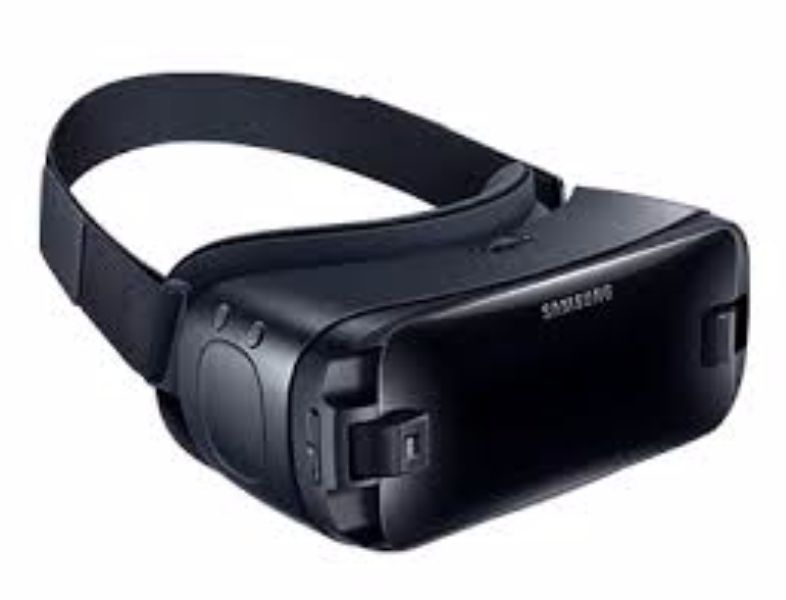}
	\caption{Samsung Gear VR\footnote{\url{http://www.samsung.com/global/galaxy/gear-vr/}}.}
	\label{fig:commercial:samsung-vr}
\end{subfigure}
\begin{subfigure}{0.3\columnwidth}
	\centering
	\includegraphics[width=\columnwidth]{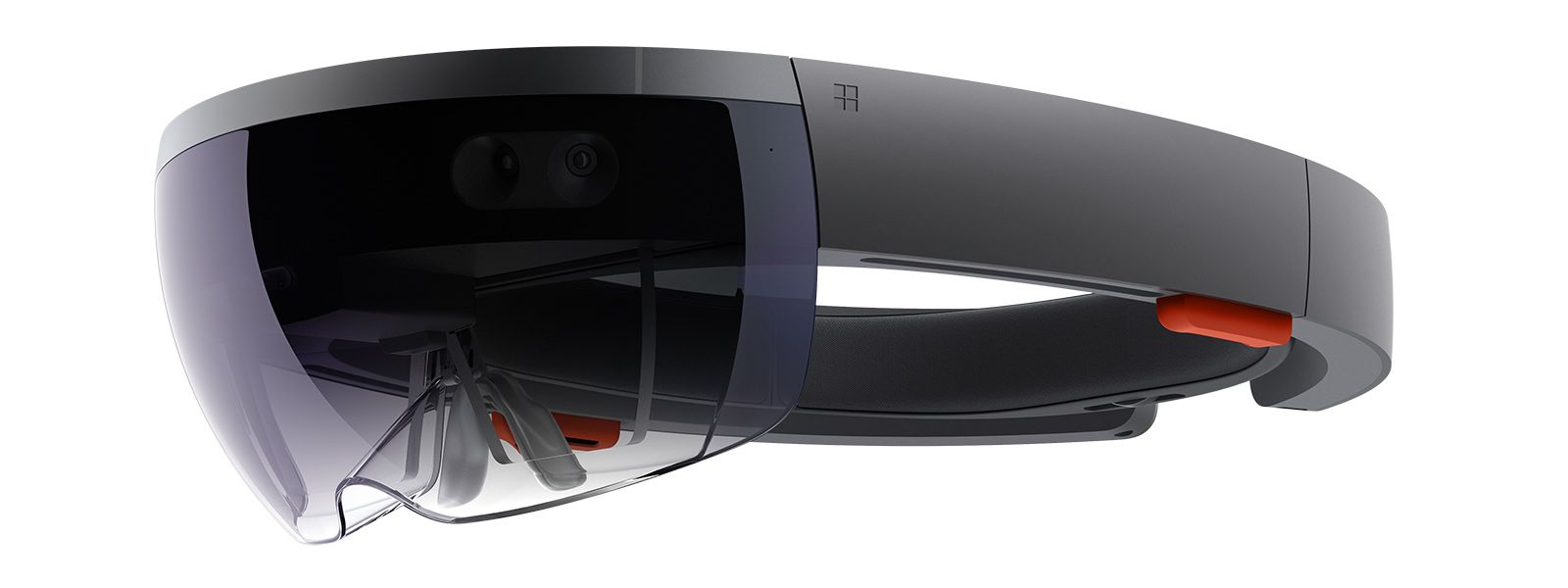}
	\caption{Microsoft Hololens\footnote{\url{https://www.microsoft.com/en-us/hololens}}.}
	\label{fig:commercial:microsoft-hololens}
\end{subfigure}
\begin{subfigure}{0.3\columnwidth}
	\centering
	\includegraphics[width=\columnwidth]{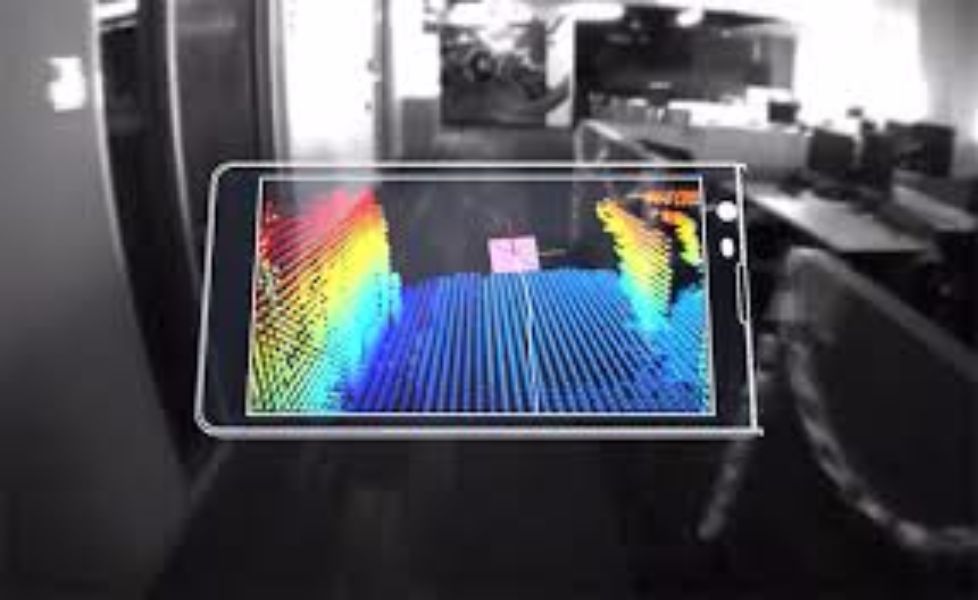}
	\caption{Google Tango project\footnote{\url{https://get.google.com/tango/}}.}
	\label{fig:commercial:google-tango}
\end{subfigure}
\begin{subfigure}{0.3\columnwidth}
	\centering
	\includegraphics[width=\columnwidth]{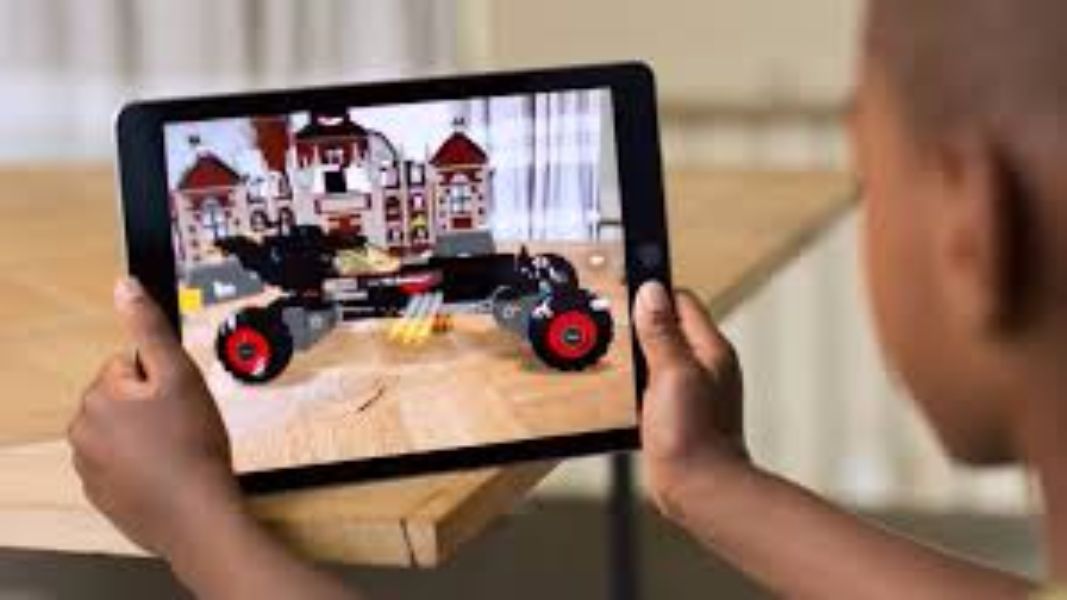}
	\caption{Apple ARKit\footnote{\url{https://developer.apple.com/arkit/}}.}
	\label{fig:commercial:apple-ar}
\end{subfigure}
	\caption{Commercial VR, AR ecosystems.}
	\label{fig:commercial} 
\end{figure}

\section{The VR and AR ecosystem} \label{sec:commercial}

The VR and AR ecosystems has been gaining more importance in recent years with the commercialization of several VR and AR devices such as Oculus Rift, HTC Vive for gaming, Google Glass, and Microsoft Hololens for AR applications (Figure\ref{fig:commercial}). Furthermore, important software and device companies are opening their frameworks to developers for AR and MAR applications (i.e., Apple ARKit\footnote{\url{https://developer.apple.com/arkit/}}).

Although many devices and technologies focus their attention on VR ecosystem, most of these solutions can be extrapolated to AR scenarios. In case of MAR applications, the compactness, weight design are important due to the mobile nature of the scenarios. Some devices such as HTC Vive and Oculus Rift provide controllers with haptic feedback (i.e., vibration) to interact and feel virtual environments. These device were designed for VR scenarios, but their features make them feasible for AR and MAR applications, due to the compactness and light weight of these devices.

\begin{figure}[t]
\centering
	\includegraphics[width=0.5\columnwidth]{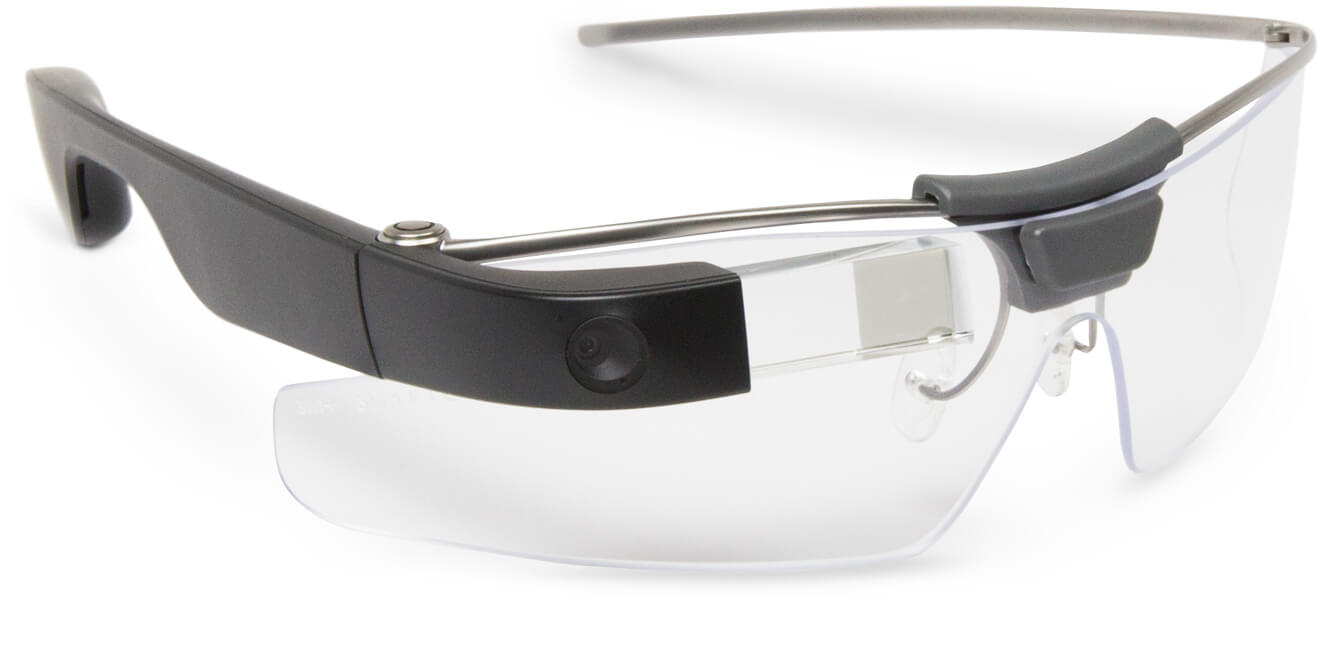}
	\caption{Google glass X, industry focused.}
	\label{fig:commercial:google-glass} 
\end{figure}

Google Glass started the revolution of MAR devices. However, due to privacy concerns and other issues the Google Glass project stopped. The new environment of this device, Google Glass 2\footnote{\url{https://x.company/glass/}} spins around enterprises such as medicine, logistics, and manufacturing, Figure~\ref{fig:commercial:google-glass}. We can realize about the high spec AR devices that the industry provides, but there is still a gap in the communication between the virtual objects and the users. Haptic feedback provides the best approach to enhance and improve the other audio-visual feedback channels, and achieve an overall better UX.

\section{Challenges and Future directions} \label{sec:challenges}

Pacchierotti et al~\cite{pacchierotti2017:wearable} present a taxonomy of current wearable haptic devices for hands and fingerprints. They also discuss the future paths and challenges of wearable haptic interfaces. The paper focuses its attention on finger and hand wearable devices. The authors describe in detail the current devices and different interaction methods, and wearable haptic feedback that has been proposed. They also enumerate the requirements and considerations needed to design wearable devices, such as comfort, impairment (i.e., mobility of the user while using the wearable device), size and weight. The authors estimate that gaming and VR are the paths to follow as the current and future gaming market will greatly benefit from haptic advances. Aside from that, robotic teleoperation is another topic which has been studied and will continue to be relevant in the haptic field. To conclude, the authors mention the possibilities for assistive and privacy aware applications. In the latter, wearable devices will enable new notifications interfaces that only the receiver will notice.

\begin{figure}[t]
  \centering
  \begin{subfigure}{0.46\columnwidth}
      \centering
      \includegraphics[width=\columnwidth]{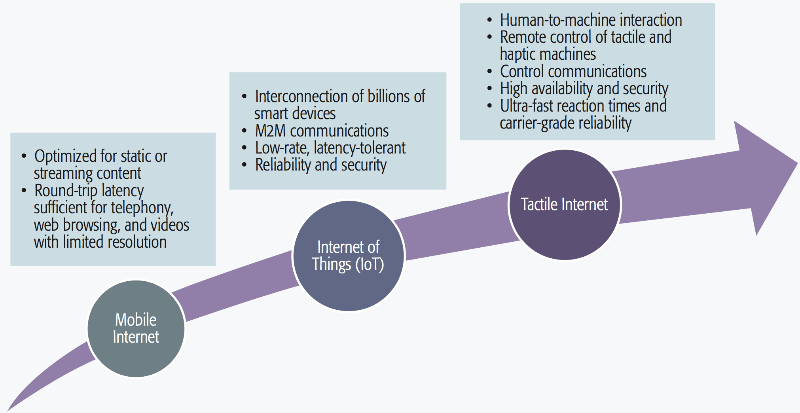}
      \caption{Revolutions.}
      \label{fig:tactile-network-evolution}
  \end{subfigure}
  \begin{subfigure}{0.46\columnwidth}
      \centering
      \includegraphics[width=\columnwidth]{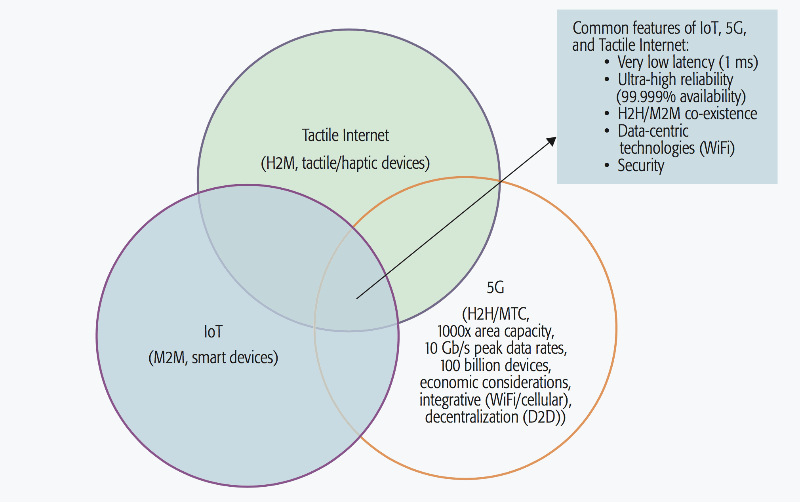}
      \caption{Network challenges.}
      \label{fig:tactile-network-challenges}
  \end{subfigure}
  \caption{Tactile Internet the next revolution, Figures by~\cite{Maier2016}}
  \label{fig:tactile-internet} 
\end{figure}

The network requirements for future MAR application feasibility needs to be tackled. Network requirements such as latency and transmission errors (e.g., teleoperation robots) are important topics to study. New advances in electronics and future wireless communications will lead to real-time interactions with our distant environments (e.g., teleoperations), Fettweis~\cite{Fettweis2014}, coined the ecosystem of the \textit{Tactile Internet}. The \textit{Tactile Internet} presents several challenges for mobile networks and also Internet's backbone such as latency and ultra-high reliability~\cite{Maier2016}~\cite{Aijaz2016}~\cite{Simsek2016}, Figure~\ref{fig:tactile-internet}. The \textit{Tactile Internet} requires \textit{1 ms} delay to achieve real-time haptic performance in scenarios such as surgery teleoperations. Pilz et al.~\cite{Pilz2016} demonstrate the implementation of wireless network towards 5G with \textit{1ms} delay requirements. In~\cite{Bachhuber2017}, the authors analyze the current video protocols to show that current technologies are not ready in terms of requirements for the \textit{Tactile Internet}. Only the delay from glass-to-glass (i.e., time between video recorded by smartphone camera until is rendered in the smartphone screen) is considerable bigger (19.18ms) than the expected \textit{1ms} for the round-trip delay of the \textit{Tactile Internet}. Therefore, the accumulated delay from the mobile network will not satisfy the latency \textit{Tactile Internet} demands. Popovski~\cite{Popovski2014} analyzes the current mechanisms to provide Ultra-Reliable Communication (URC) in 5G wireless systems. URC will bring high reliable connectivity for the next generation applications such as vehicular-to-vehicular communications, \textit{Tactile Internet}, and sensor networks over 5G cellular networks. However, the high reliability capabilities can affect the stringent latency requirements of the mentioned services. Besides the number of users contribute to this trade-off latency--high-reliability.   

Furthermore, the demanding requirement of \textit{1 ms} latency is not only limited by  mobile and backbone networks but from sensing devices. The authors in~\cite{Steinbach2012}~\cite{Wildenbeest2013} analyze multimodal integration approaches to aggregate different haptic human sensing information to the network to achieve the latency requirements and not affect the reliability of the ecosystem. Steinbach et al~\cite{Steinbach2012} study how human perceptions work and how human brain combines the sensory information. The authors aim to integrate different sensory information without decreasing the percept effectiveness and accuracy. Besides, they study delay effects in haptic sensing tasks. In~\cite{Wildenbeest2013}, the authors study the overall performance for teleoperation tasks according to different haptic feedbacks. 

In order to design haptic feedback devices and its mechanical interfaces, there is an approach that assumes passivity for the haptic device and stable human interaction with the device. The passivity condition formulated by Colgate and Schenkel~\cite{colgate1994:passivity} utilizes this fact. The authors in~\cite{hulin2014:passivity} extend this approach to virtual walls (1DOF). The device used for the analysis consists of discrete-time spring-damper system. They demonstrate why passivity is conservative for haptic systems in terms of stability.

Muller et al.~\cite{muller2016:immersive} propose a collaborative mixed reality interface. The immerse exploration can help to explore and understand health related data. Collaborative techniques can enhance of current AR/MAR application usability.

\textit{Deadband} compression techniques transmit new haptic data only when the stimuli can be perceived by user (based on Just Noticeable Difference, JND). In~\cite{Tirmizi2016}, the authors extend the work on deadband approaches for cutaneous haptic data. The proposed device to measure the JND consists of an Omega 3 interface and a custom 3DOF cutaneous device. The compression technique for cutaneous haptic feedback uses the JND as the perceptual threshold for the compression algorithm. The result shows an average reduction of around 60\%. The benefits of deadband compression algorithms are plausible and the implementation of the JND threshold algorithm feasible. Difficulties can appear in the perception measures, as it is a personal characteristic that can  vary between users.

The design and development of AR/MAR applications need to consider other populations such as the elderly, or visually impaired to not leave them behind in the forthcoming AR/MAR era. Liang~\cite{liang2016:design} presents the eight elements that need to be considered in an AR architecture such as user, interaction, device, virtual content, real content, transmission, server and physical world. The author also describes five preliminary design principles of AR systems: changeability, synchronization, partial one to one, antecedent and hidden reality. The authors identify an AR pillbox as an example of AR for the aging population. Bach-y-Rita et al.~\cite{bach1969:vision} present one of the first works to substitute vision by tactile sensing. Pawluk et al. describe the relevant issues on designing haptic assistive technologies for the visually impaired. The first issue using tactile mechanical feedback as a substitution for vision is the low space resolution of the tactile sensing. Besides the field-of-view for vision is considerably smaller for touch. Another approach uses eletrotactile displays~\cite{spelmezan2016:sparkle} to simulate a mechanical interaction. However, it suffers from similar limitations. One well known example of vision substitution using eletroctactile feedback is the tongue-based feedback device proposed by Bach-y-Rita et al.~\cite{bach1998:form} and Lozano et al.~\cite{lozano2009:electrotactile}. Furthermore, performance of assistive technologies differs greatly from laboratories (i.e., testing environments) and the real world. The authors also mention the effect of affective touch experiences, and their importance in visual substitution approaches. There are several areas of haptic assistive technology such as Braille devices (i.e., piezolectric pin arrays, virtual Braille displays), tactile rendering techniques for graphics and mobility devices (i.e., smart canes). One of the main differences in the way we understand information is that it is not possible to map the visual scene into the tactile sensory skin.  The visually impaired population should be considered not only for assistive technologies but for the AR/MAR ecosystem (i.e., not only visual feedback).

The addition of haptic feedback to the learning process has been studied in several works. In~\cite{han2011:incorporating}, the authors include kinesthetic feedback in physic simulations to help students understand forces involved in gears mechanical behavior. Okamura et al.~\cite{okamura2002:feeling}, follow a similar approach using a force feedback joystick to teaching dynamic systems. Minogue et al.~\cite{minogue2006:haptics}, analyze the understanding improvements using touch roles in the cognitive-learning process, and efficacy in haptic augmentation. Another important field that can be enhanced by haptic feedback is medical training~\cite{coles2011:role}. Haptic feedback can facilitate the learning process and also simulation training with more realistic scenarios. In~\cite{richard2006:multi}, the authors realize a survey about olfactory feedback, haptic feedbacks and immersive applications applied in teaching and learning; for example, to teach the abstract concept of the Bohr atomic model. Haptic feedbacks can also provide novel ways of passive learning. For example, the use of haptic actuators can play a similar role as learning by seeing: as learning by haptic feeling. In~\cite{Seim2015}, the authors propose a passive haptic learning (PHL) stimulation method to teach piano pieces. The authors use a pair of wearable gloves with vibroactuators on the back of each finger. They study the efficiency of this PHL approach together with audio and only vibration feedback. The results show that vibration and vibration + audio techniques are useful for learning and retention, as they offer better performance than without them (i.e., only audio). These innovative techniques to improve the learning process are worth studying and can be a new part of development of tools together with MAR/AR applications.

\section{Conclusion} \label{sec:conclusion}

In this survey, we depict the state-of-the-art of several haptic devices and their capabilities as wearables in MAR ecosystem. Furthermore, we classify the haptic feedback devices by their sensory nature and their design characteristics, such as mid-air, and exoskeleton. We start with a brief description of the main features of haptic devices and the importance of audio and visual as non-haptic devices in enhancing the UX and improving the overall interaction performance. We analyze the main characteristics of the proposed devices, and their applicability as wearables for MAR applications.
Although there are many works and commercial products, an affordable, portable and simple approach for haptic wearable devices is still missing. Moreover, the fidelity of these devices is limited to one scenario such as surface/texture rendering, grasping, or pushing. The combination of more haptic devices to achieve better feedback has been done by several authors but the size, or difficult implementation hinders their deployment in mobile environments, where the scenarios and circumstances surrounding the user can change.
With this work we aim to provide a better understanding of mechanisms, challenges and future possibilities of haptic feedback in the MAR field.


\bibliographystyle{IEEEtran}
\bibliography{survey_haptics}

\end{document}